\documentclass[aps,prd,preprint,eqsecnum,floats,epsf,superscriptaddress,nofootinbib]{revtex4}
\usepackage{epsfig}
\usepackage{graphicx}
\usepackage{extarrows}
\usepackage{subfigure}
\usepackage{color}

\def\ol{\overline}
\def\ov{\overline}
\def\be{\begin{eqnarray}}
\def\en{\end{eqnarray}}
\def\vp{\varepsilon}
\def\non{\nonumber}
\def\CP{{\it CP}~}

\def\la{\langle}
\def\ra{\rangle}

\def\B{{\cal B}}

\def\lsim{ {\ \lower-1.2pt\vbox{\hbox{\rlap{$<$}\lower5pt\vbox{\hbox{$\sim$}
}}}\ } }
\def\gsim{ {\ \lower-1.2pt\vbox{\hbox{\rlap{$>$}\lower5pt\vbox{\hbox{$\sim$}
}}}\ } }

\begin{document}

\vskip 1.0 cm

\title{Updated analysis of $D\to PP, V\!P$ and $VV$ decays:\\
Implications for $K_S^0-K_L^0$ asymmetries and $D^0$-$\ov D^0$ mixing}

\author{Hai-Yang Cheng}
\affiliation{Institute of Physics, Academia Sinica, Taipei, Taiwan 11529, ROC}

\author{Cheng-Wei Chiang}
\affiliation{Department of Physics and Center for Theoretical Physics, National Taiwan University, Taipei, Taiwan 10617, ROC}
\affiliation{Physics Division, National Center for Theoretical Sciences, Taipei, Taiwan 10617, ROC}

\medskip
\begin{abstract}
\bigskip

An updated analysis of the two-body $D\to PP, V\!P$ and $VV$ decays within the framework of the topological diagram approach is performed.  A global fit to the Cabibbo-favored (CF) modes in the $V\!P$ sector gives many solutions with similarly small local minima in $\chi^2$.  The solution degeneracy is lifted once we use them to predict for the singly Cabibbo-suppressed (SCS) modes. Topological amplitudes are extracted for the $\eta-\eta'$ mixing angles $\phi=40.4^\circ$ and $43.5^\circ$.
The $K_S^0-K_L^0$ asymmetries in $D\to K_{S,L}^0M$ decays denoted by $R(D,M)$ are studied. While the predicted $R(D^0,P)$  for $P=\pi^0, \eta$ and $\eta'$ agree with experiment, the calculated $R(D^+,\pi^+)$, $R(D_s^+, K^+)$, $R(D^0,\omega)$ and $R(D^0,\phi)$ deviate from the data. We conjecture that the relative phase between the topological amplitudes $(C+A)$ and $(T+C)$ should be slightly smaller than $90^\circ$ in order to explain the first two discrepancies and that additional singlet contributions due to the SU(3)-singlet nature of $\omega$ and $\phi$ are needed to account for the last two. For doubly Cabibbo-suppressed (DCS) $D\to V\!P$ decays, their topological amplitudes (double-primed) cannot be all the same as the corresponding ones in the CF modes.
The assumption of $E_{V,P}''=E_{V,P}$ for the $W$-exchange amplitude leads to some inconsistencies with the experiment. Through the measured relative phases between CF and DCS channels,
the relations of $E_{V,P}''$ with $E_{V,P}$ are determined.
Long-distance contributions to the $D^0$-$\ov D^0$ mixing parameter $y$ are evaluated in the exclusive approach.  In particular, we focus on $D\to PP$ and $V\!P$ decays where $y$ can be reliably estimated.  We conclude that $y_{_{P\!P}}\sim (0.110\pm 0.011)\%$ and the lower bound on $y_{_{V\!P}}$ is $(0.220\pm 0.071)\%$. It is thus conceivable that at least half of the mixing parameter $y$ can be accounted for by the two-body $PP$ and $V\!P$ modes. The main uncertainties arise from the yet-to-be-measured DCS channels and their phases relative to the CF ones.

\end{abstract}


\maketitle
\small

\newpage

\section{Introduction \label{sec:intro}}

Contrary to the bottom sector where the physics of two-body nonleptonic $B$ decays can be formulated in a QCD-inspired approach, a theoretical description of the underlying mechanism for exclusive hadronic $D$ decays based on QCD is still absent today.  This has to do with the mass of the charm quark, of order 1.5 GeV.  It is not heavy enough to allow for a sensible heavy quark mass expansion and not light enough for an application of chiral perturbation theory.  Nevertheless, a model-independent analysis of charm decays based on the topological diagram approach (TDA) is achievable. In this approach, the topological diagrams are classified according to the topologies of weak interactions with strong interaction effects at all orders implicitly taken into account.

Analyses of $D\to P\!P$ and $V\!P$ decays have been performed in Refs.~\cite{Cheng:2010,Cheng:2012a,Cheng:2012b,Cheng:2016,Cheng:2019,Cheng:2021} within the framework of TDA, where $P$ and $V$ denote respectively pseudoscalar and vector mesons. In this work, we shall perform an updated analysis for the following reasons: (i) In previous analyses, we have set $\Gamma(\ov K^0)=2\Gamma(K_S^0)$  for decay modes involving a neutral $K_S^0$.  But this relation can be invalidated by the interference between Cabibbo-favored (CF) and doubly Cabibbo-suppressed (DCS) amplitudes. (ii) Branching fraction measurements of several CF and singly Cabibbo-suppressed (SCS) $D\to V\!P$ modes have been significantly improved in recent years. These new data provide valuable information and modify the sizes and phases of the relevant topological amplitudes. (iii) DCS decays were not carefully studied in the previous analyses, particularly due to the lack of sufficiently precise data in the $V\!P$ channels. We shall examine whether the topological amplitudes in the DCS sector are the same as those in the CF decays. We will calculate the $K_S^0-K_L^0$ asymmetries in $D\to K_{S,L}^0M$ decays with $M=P,V$ and compare them with the experiment. This will provide information on the DCS topological amplitudes. (iv) Thanks to BESIII, many new data on $D_s^+$ and $D^+$ decays to $V\!V$ became available in the past few years. It turns out that several $D\to V\!V$ modes are dominated by the $D$-wave amplitude and some dominated by the $P$-wave, contrary to the na{\"i}ve expectation of $S$-wave dominance. Moreover, the decay $D^0\to \omega\phi$ has been observed by BESIII to be transversely polarized with negligible longitudinal polarization~\cite{BESIII:D0omegaphi}. In the end, there are many puzzles in the $V\!V$ sector that need to be resolved. (v) One of our goals is to evaluate the $D^0$-$\ov D^0$ mixing parameter which we are going to elaborate on below.

The $D^0$-$\ov D^0$ mixing occurs because the mass eigenstates $D_{1,2}$ are not the same as the flavor eigenstates $D^0$ and $\ov D^0$. $D$ mixing is conventionally described by the two parameters $x\equiv \Delta m/\Gamma$ and $y\equiv \Delta\Gamma/2\Gamma$, where $\Delta m=m_1-m_2$ defined to be positive and $\Delta\Gamma=\Gamma_1-\Gamma_2$. Evidence for the $D^0$-$\ov D^0$ mixing has been established and the current world averages for the {\it CP} allowed case are~\cite{HFLAV}:
\be \label{eq:WA}
x=(0.409^{+0.048}_{-0.049})\%, \qquad y=(0.615^{+0.056}_{-0.055})\%.
\en
The absence of mixing, namely $x=y=0$, is excluded at $11.5\sigma$. Very recently, LHCb has made a model-independent measurement of charm mixing parameters in $B\to D^0(\to K_S^0{\pi}^+{\pi}^-){\mu}^-\bar{\nu}_{\mu}X$ decays and obtained
the results~\cite{LHCb:Dmixing}
\be
x=(0.401\pm0.049)\%, \qquad y=(0.55\pm0.13)\%,
\en
which are consistent with current averages given in Eq. (\ref{eq:WA}).

Early calculations of short-distance contributions to the $D$ mixing parameters already
indicated that both $x$ and $y$ were very small, of order $10^{-6}$, due to the Glashow-Iliopoulos-Maiani suppression~\cite{Cheng:1982,Datta}. This implies that the observed $D^0$-$\ov D^0$ mixing is dominated by long-distance processes. Indeed, it is well-known that charm physics is governed by non-perturbative effects. In the literature, the $D^0$-$\ov D^0$ mixing is usually studied in three different approaches: inclusive, exclusive, and dispersive. In the inclusive approach, the mixing parameters are systematically investigated based on heavy quark expansion (HQE) dictated by the parameter $1/m_c$~\cite{Golowich:2005,Lenz:2010,Lenz:2017,Lenz:2020,Umeeda} (for a review, see Ref.~\cite{Lenz:2021}). Since the lifetime of $D_s^+$ and the ratio $\tau(D^+)/\tau(D^0)$ have been found to be in good agreement with experiment within the framework of HQE~\cite{Lenz:2022}, it is natural to expect that this approach might also be viable for the mixing parameters. However, it turns out that the suppression of short-distance contributions cannot be alleviated in HQE.

Contrary to the HQE approach at the quark level, long-distance contributions from the intermediate hadronic states are summed over in the exclusive approach~\cite{Wolfenstein:1985,Donoghue:1986,Colangelo,Kaeding,Falk:y,Cheng:Dmixing,Gronau,Jiang}. As pointed out in Ref.~\cite{Falk:y}, the mixing parameters $x$ and $y$ vanish in the flavor SU(3) limit. In general, there are large cancellations of CF and DCS decays with the contributions from SCS
decays. The cancellation will be perfect in the limit of SU(3) symmetry, and the $D^0$-$\ov D^0$ mixing would occur only at the second order in SU(3) breaking.  For example, contributions from SCS decays $D^0\to\pi^+\pi^-, K^+K^-$ are canceled by contributions from the CF $D^0\to K^-\pi^+$ decay and the DCS $D^0\to K^+\pi^-$ decay. Since the intermediate states in this case are related by U-spin symmetry, it has been shown in Ref.~\cite{Gronau} that the $D^0$-$\ov D^0$ mixing in the Standard Model occurs only at the second order in U-spin breaking.

A dispersive relation between $x$ and $y$ has been derived in Ref.~\cite{Falk:x} in the heavy quark limit
\be
\Delta m=-{1\over 2\pi}{\cal P}\int_{2m_\pi}^\infty dE\left[ {\Delta\Gamma(E)\over E-m_D}+{\cal O}
\left({\Lambda_{\rm QCD}\over E}\right) \right].
\en
For a given model of $\Delta\Gamma(E)$ or $y(E)$, it is conceivable to have $x$ comparable to $y$ in magnitude.
Writing
\be
\int^\Lambda_0 ds'{y(s')\over s-s'}=\pi x(s)-\int^\infty_\Lambda ds' {y(s')\over s-s'}
\en
and choosing the scale $\Lambda$ large enough to justify the perturbative calculation of $y$
on the right-hand side, and below the $b$ quark threshold to avoid the $b$ quark contribution to the left-hand side, the authors of Ref. \cite{Li:inverse} showed that the study of the $D$ meson mixing was converted into an inverse problem: the mixing parameters at low masses are solved as source distributions, which produce the potential observed at high masses. The analysis is further improved in Ref.~\cite{Li:2022} in which SU(3) breaking is introduced through physical thresholds of different $D$ meson decay channels.

In this work, we shall focus on the exclusive scenario for the $D^0$-$\ov D^0$ mixing parameters.
Since the two-body $D^0\to PP$ decays and quasi-two-body decays such as $D^0\to V\!P, V\!V, SP, SV, AP, AV$, $TP,TV$ account for about 3/4 of the total hadronic rates, where $S$, $A$ and $T$ denote respectively the scalar, axial-vector and tensor mesons, it is arguable that these two-body and quasi-two-body channels dominate and provide a good estimate of the mixing parameters. Data on two-body $D\to PP, V\!P$ decays have been accumulated in the past few years with substantially improved precision~\cite{PDG}. For example, the measurements of all $D^0\to PP$ channels are available except for three of the DCS modes. Hence, in principle, one can estimate $y_{PP}$ directly from the data. The DCS decays are related to the CF ones through the relations such as $\B(D^0\to K^0\eta^{(')})=\tan^4\theta_C\B(D^0\to \ov K^0\eta^{(')})$, where $\theta_C$ is the Cabibbo angle.  However, we have applied the TDA in Ref.~\cite{Cheng:Dmixing} to estimate the $D$ mixing parameters for the following reasons: (i) to predict the branching fractions of yet-to-be-measured SCS and DCS modes in the $D\to V\!P$ decays, (ii) to understand SU(3) breaking effects in the SCS and DCS modes, (iii) to see explicitly the vanishing mixing parameters in the SU(3) limit, and (iv) to reduce the uncertainties in the estimate of the parameter $y$. A similar study based on the so-called factorization-assisted topological-amplitude (FAT) approach has been carried out in Ref.~\cite{Jiang}.

The layout of the present paper is as follows. The analysis of $D\to P\!P$ decays within the framework of TDA is presented in Sec.~\ref{sec:DtoPP} for $\phi=40.4^\circ$ and $43.5^\circ$ with $\phi$ being the $\eta-\eta'$ mixing angle. As for $D\to V\!P$ decays, its analysis is much more complicated. We shall discuss the CF, SCS and DCS decays separately and consider the $K_S^0-K_L^0$ asymmetries in $D\to K_{S,L}^0V$ channels in Sec.~\ref{sec:DtoVP}. Sec.~\ref{sec:DtoVV} is devoted to the discussions of $D\to V\!V$ decays. We evaluate the $D^0$-$\ov D^0$ mixing parameter $y$ in Sec.~\ref{sec:y} with a focus on the contributions from the $P\!P$ and $V\!P$ sectors. Sec.~\ref{sec:Conclusions} comes to our conclusions.

\section{$D\to P\!P$ decays \label{sec:DtoPP}}

\subsubsection{Cabibbo-favored $D\to P\!P$ decays}

It was established some time ago that a least model-dependent analysis of heavy meson decays could be carried out in the TDA~\cite{Chau,CC86,CC87}. For the present purposes, it suffices to consider tree amplitudes: color-allowed tree amplitude $T$; color-suppressed tree amplitude $C$; $W$-exchange amplitude $E$ and $W$-annihilation amplitude $A$.  The topological amplitudes for CF $D\to P\!P$ decays~\cite{PDG} are shown in Table~\ref{tab:AmpDtoPP:CF} where $\phi$ is the $\eta-\eta'$ mixing angle defined in the flavor basis
\be
 \begin{pmatrix} \eta \cr \eta' \end{pmatrix}
 = \begin{pmatrix} \cos\phi & -\sin\phi \cr \sin\phi & \cos\phi\cr \end{pmatrix}
 \begin{pmatrix} \eta_q \cr \eta_s \end{pmatrix},
\en
with $\eta_q={1\over\sqrt{2}}(u\bar u+d\bar d)$ and $\eta_s=s\bar s$.
For the $\eta-\eta'$ mixing angle, an early study gave $\phi=(39.3\pm1.0)^\circ$~\cite{Feldmann}. This mixing angle has been measured by KLOE to be $\phi=(40.4\pm0.6)^\circ$~\cite{KLOE}.  We have previously followed the LHCb measurement of $\phi=(43.5^{+1.4}_{-1.3})^\circ$~\cite{LHCb:etaetap} to fix $\phi$ to be $43.5^\circ$.
Recent precision measurements of $D_s^+\to \eta^{(')}e^+\nu_e$  and $D_s^+\to \eta^{(')}\mu^+\nu_\mu$ by BESIII yield $\phi=(40.0\pm2.0\pm0.6)^\circ$~\cite{BESIII:etaetap_a} and $\phi=(40.2\pm2.1\pm0.7)^\circ$~\cite{BESIII:etaetap_b}, respectively.
In this work, we will study the cases for $\phi=40.4^\circ$ and $43.5^\circ$.  Though these two choices differ by only a few degrees, they do produce observable differences in fitting the data. We shall see that the former mixing angle is preferred by the $D\to P\!P$ data, whereas the latter is slightly favored by the $D\to V\!P$ data.

\begin{table}[t!]
\caption{Topological amplitude representation and branching fractions for the CF $D\to PP$ decays.  Data are taken from Ref.~\cite{PDG}. Here $\lambda_{sd}\equiv V_{cs}^*V_{ud}$.
  \label{tab:AmpDtoPP:CF}  }
\medskip
\scriptsize{
\begin{ruledtabular}
\begin{tabular}{l  c c  l c c  }
  Mode~~~ & ~~~~Representation~~~~~~ & $\B_{\rm expt}$ (\%) & ~~~Mode & Representation & $\B_{\rm expt}$ (\%) \\
\hline
  $D^0\to K^{-}\pi^+$ & $\lambda_{sd}(T+E)$ & $3.947\pm0.030$ & ~~~$D^0\to \ov K^{0}\eta$~~~ & $\lambda_{sd}[ {1\over\sqrt{2}}(C+E)\cos\phi-E\sin\phi]$ & $0.958\pm0.0020$ \\
  $D^0\to \ov K^{0}\pi^-$ & ${1\over\sqrt{2}}\lambda_{sd}(C-E)$ & $2.311\pm0.036$ &~~~$D^0\to \ov K^{0}\eta'$
  & $\lambda_{sd}[ {1\over\sqrt{2}}(C+E)\cos\phi-E\sin\phi]$ & $1.773\pm0.047$ \\
  $D^+\to \ov K^{0}\pi^+$ & $\lambda_{sd}(T+C)$ & $3.067\pm0.053$ &\\
  $D_s^+\to \ov K^0 K^+$ & $\lambda_{sd}(C+A)$ & $2.202\pm0.060$ &~~~ $D_s^+\to\pi^+\eta$ &  $\lambda_{sd}[\sqrt{2}A\cos\phi-T\sin\phi]$ & $1.68\pm0.09$ \\
  $D_s^+\to \pi^+\pi^0$ & 0 & $<0.012$ &~~~ $D_s^+\to\pi^+\eta'$ &  $\lambda_{sd}[\sqrt{2}A\sin\phi+T\cos\phi]$ & $3.94\pm0.25$ \\
\end{tabular}
\end{ruledtabular}}
\end{table}
%

The topological amplitudes for the Cabibbo-favored $D\to P\!P$ decays~\cite{PDG} are shown in Table~\ref{tab:AmpDtoPP:CF}. For the CF decay modes involving a neutral kaon $K_S^0$ or $K_L^0$, it was customary to use the relation $\Gamma(\ov K^0)=2\Gamma(K_S^0)$. However, this relation can be invalidated by the interference between CF and DCS amplitudes. Using the phase convention that $K_S^0={1\over\sqrt{2}}(K^0-\ov K^0)$ and $K_L^0={1\over\sqrt{2}}(K^0+\ov K^0)$ in the absence of $C\!P$ violation, we have
\be
\begin{split}
A(D\to K_S^0 M) &= -{1\over\sqrt{2}}[A(D\to \ov K^0M)-A(D\to K^0 M)],
\\
A(D\to K_L^0 M) &= {1\over\sqrt{2}}[A(D\to \ov K^0M)+A(D\to K^0 M)],
\end{split}
\en
where $M=P$ or $V$.
Consequently, $\B(D\to K_S^0 M)+\B(D\to K_L^0 M)=\B(D\to \ov K^0 M)+\B(D\to K^0 M)$. Hence,
we shall use $\B(D\to \ov K^0 M)\cong \B(D\to K_S^0 M)+\B(D\to K_L^0 M)$ which is valid to a good approximation. For example, using the measured branching fractions of $D^0\to K_{S,L}^0\pi^0$ given in Table~\ref{tab:PPKLKS} below, we shall get $\B(D^0\to\ov K^0\pi^0)=(2.311\pm0.036)\%$. Of course, one can also perform a fit to the data of $D\to K_S^0P$ and/or $D\to K_L^0 P$ instead of $D\to \ov K^0P$, assuming that the DCS (double-primed) amplitudes are the same as the CF (unprimed) ones. We shall see later that it is more convenient to consider the CF $D^0\to \ov K^0P$ and DCS $D^0\to K^0P$ decays in order to compute the $D^0$-$\ov D^0$ mixing parameter $y$.

\begin{table}[t]
\caption{Topological amplitudes extracted from the CF $D\to PP$ decays in units of $10^{-6}$~GeV.
The color-allowed amplitude $T$ is taken to be real.
Previous fits obtained in 2010~\cite{Cheng:2010} and 2019~\cite{Cheng:2019} are also listed for comparison.
  \label{tab:AmpPP}}
  \medskip
\footnotesize{
\begin{ruledtabular}
\begin{tabular}{ l c  c c c c c c c}
 Year &  $T$ & $C$ & $E$ & $A$ \\
\hline
2010  & $3.14\pm0.06$\footnotemark[1] & $(2.61\pm0.08)e^{-i(152\pm1)^\circ}$ & $(1.53^{+0.07}_{-0.08})e^{i(122\pm2)^\circ}$ & $(0.39^{+0.13}_{-0.09})e^{i(31^{+20}_{-33})^\circ}$ \\
2019  & $3.113\pm0.011$\footnotemark[2] & $(2.767\pm0.029)e^{-i(151.3\pm0.3)^\circ}$ & $(1.48\pm0.04)e^{i(120.9\pm0.4)^\circ}$ & $(0.55\pm0.03)e^{i(23^{+~7}_{-10})^\circ}$ \\
2023a & $3.134\pm0.010$\footnotemark[1] & $(2.584\pm0.014)e^{-i(151.9\pm0.3)^\circ}$ & $(1.472\pm0.024)e^{i(121.7\pm0.4)^\circ}$ & $(0.394\pm0.020)e^{i(14.1^{+11.0}_{-~8.5})^\circ}$\\
2023b & $3.175\pm0.010$\footnotemark[2] & $(2.711\pm0.014)e^{-i(152.1\pm0.3)^\circ}$ & $(1.350\pm0.025)e^{i(123.8\pm0.4)^\circ}$ & $(0.541\pm0.021)e^{i(9.4^{+6.5}_{-5.2})^\circ}$
\end{tabular}
\footnotetext[1]{For $\theta=40.4^\circ$.} \footnotetext[2]{For $\theta=43.5^\circ$.}
\end{ruledtabular}}
\end{table}

It is clear from Table~\ref{tab:AmpDtoPP:CF} that we have 8 data points for 7 unknown parameters.
Hence, the topological amplitudes $T$, $C$, $E$ and $A$ can be extracted from the CF $D\to P\!P$ decays through a $\chi^2$ fit, as shown in Table~\ref{tab:AmpPP} for $\phi=40.4^\circ$ and $43.5^\circ$, respectively. The fitted $\chi^2$ value almost vanishes with the fit quality of $99.6\%$ for $\phi=40.4^\circ$ and is 1.46 per degree of freedom with the fit quality of 22.7\% for $\phi=43.5^\circ$.  Previous fits obtained in 2010~\cite{Cheng:2010} and 2019~\cite{Cheng:2019} are also listed in Table~\ref{tab:AmpPP} for comparison. We see that the errors in $T$, $C$, $E$, and $A$ are substantially reduced, especially for the annihilation amplitude $A$, thanks to the improved data precision from the Particle Data Group (PDG)~\cite{PDG}.

We have noticed before~\cite{Cheng:2019} that since we fit only the observed branching fractions, the results will be the same if all the strong phases are subject to a simultaneous sign flip, resulting in a two-fold ambiguity.  Throughout this paper, we only present one of them.  Presumably, such a degeneracy in strong phases can be resolved by measurements of sufficiently many \CP asymmetries. For example, a measurement of direct \CP asymmetry in $D^0\to K_S^0K_S^0$ will allow us to resolve the discrete phase ambiguity~\cite{Cheng:2019}.

We see in Table~\ref{tab:AmpPP} that the topological amplitudes respect the hierarchical pattern $|T|>|C|>|E|>|A|$. The phase between $C$ and $T$ is $150^\circ$, not far from the expectation of $180^\circ$ from na{\"i}ve factorization. The $W$-exchange amplitude $E$ is sizable with a large phase of order $120^\circ$. This implies the importance of $1/m_c$ power corrections as the short-distance contributions to $E$ are helicity suppressed. Notice that the $W$-annihilation amplitude is smaller than the $W$-exchange amplitude. Under na{\"i}ve factorization, all the predicted topological amplitudes except $T$ are too small compared to the values extracted from the data, implying that topological amplitudes $C$, $E$ and $A$ are dominated by long-distance, nonfactorizable effects.

\subsubsection{Singly and doubly Cabibbo-suppressed $D\to P\!P$ decays}

We follow the conventional practice to denote the primed amplitudes for SCS modes and double-primed amplitudes for DCS decays. In the flavor SU(3) limit, primed and unprimed amplitudes should be the same.  It is known that there exists significant SU(3) breaking in some of the SCS modes from the symmetry limit. For example, the rate of $D^0\to K^+K^-$ is larger than that of $D^0\to \pi^+\pi^-$ by a factor of $2.8$~\cite{PDG}, while the magnitudes of their decay amplitudes should be the same in the SU(3) limit. The observation of the $D^0\to K^0\ov K^0$ decay indicates that SU(3) symmetry must also be broken in the topological amplitude $E$. Indeed, as explained in detail in Ref.~\cite{Cheng:2019}, the large rate disparity between $K^+K^-$ and $\pi^+\pi^-$ cannot rely solely on the nominal SU(3) breaking in the tree or $W$-exchange amplitude.

We note in passing that Ref.~\cite{Cheng:2019} also studied the possibility of explaining the $K^+K^-$ and $\pi^+\pi^-$ rate difference through the penguin mechanism, as proposed, e.g., in Ref.~\cite{Brod:2012ud}.  This would require a huge $\Delta P$ comparable to or even larger than $T$ in size, where $\Delta P$ was dominated by the difference of $s$- and $d$-quark penguin contractions of 4-quark tree operators.  However, we had estimated their ratio to be only of ${\cal O}(0.01)$.  We therefore do not include penguin amplitudes in the current analysis because they have negligible contributions to the branching ratios, though they are crucial in CP asymmetry analyses.

SU(3)-breaking effects in the topological amplitudes $T'$ and $C'$ can be estimated in the factorization approach as the topological unprimed amplitudes extracted from the CF $D\to \overline K\pi$ decays have the expressions
\be
\begin{split}
\label{eq:T,C}
 T &= {G_F\over
 \sqrt{2}}a_1(K\pi)\,f_\pi(m_D^2-m_K^2)F_0^{DK}(m_\pi^2),
 \\
 C &= {G_F\over
 \sqrt{2}}a_2(K\pi)\,f_K(m_D^2-m_\pi^2)F_0^{D\pi}(m_K^2).
\end{split}
\en
SU(3) breaking effects in the $T'$ and $C'$ amplitudes of SCS modes are then addressed by comparing them with the factorizable amplitudes given by Eq.~(\ref{eq:T,C})~\cite{Cheng:2012b}. For example, we found $|T_{_{K\!K}}/T| = 1.269$ and $|T_{\pi\pi}/T| = 0.964$ in Ref.~\cite{Cheng:2019}.

\begin{table}[tp!]
\caption{Branching fractions of the CF and SCS $D\to PP$ decays in units of $10^{-2}$ and $10^{-3}$, respectively.  Experimental branching fractions are taken from Ref.~\cite{PDG}. Theory predictions are based on the topological amplitude sets PPa and PPb denoted by 2023a and 2023b, respectively in Table~\ref{tab:AmpPP} corresponding to $\phi=40.4^\circ$ and $43.5^\circ$. SU(3) breaking effects in SCS decays have been taken into account (see Table~I of Ref.~\cite{Cheng:2019}).
  \label{tab:BRPP}  }
\medskip
\scriptsize{
\begin{ruledtabular}
\begin{tabular}{l  c c c  l c c c }
  Mode~~~ & ${\cal B}_{\rm expt}$~~ &  ${\cal B}_{\rm theo}$ (PPa) &  ${\cal B}_{\rm theo}$ (PPb) & Mode~~~ & ${\cal B}_{\rm expt}$~~ &  ${\cal B}_{\rm theo}$ (PPa)  &  ${\cal B}_{\rm theo}$ (PPb)
  \\
\hline
  $D^0\to K^-\pi^+$ & $3.947\pm0.030$ & $3.947\pm0.063$ & $3.943\pm0.067$ &  $D^+\to\ov K^0 \pi^+$ & $3.067\pm0.053$   & $3.067\pm0.048$ & $3.062\pm0.049$ \\
  $D^0\to \ov K^0\pi^0$ & $2.311\pm0.036$  & $2.311 \pm 0.034$  & $2.322 \pm 0.036$ & $D_s^+\to\ov K^0K^+$ & $2.920\pm0.060$   &  $2.920 \pm 0.048$  &  $2.922 \pm 0.135$ \\
  $D^0\to \ov K^0\eta$ & $0.958\pm0.020$  & $0.958 \pm 0.004$  & $0.951 \pm 0.027$ & $D_s^+\to \pi^+\eta$  & $1.68\pm0.09$ & $1.68\pm0.14$ & $1.72\pm0.11$ \\
  $D^0\to \ov K^0\eta'$ & $1.773\pm0.047$  & $1.773 \pm 0.044$  & $1.764 \pm 0.045$ & $D_s^+\to \pi^+\eta'$  & $3.94\pm0.25$  & $3.94\pm0.17$   & $4.19\pm0.15$ \\
\hline
  $D^0\to \pi^+\pi^-$ & $1.454\pm0.024$ & $1.454\pm0.021$ & $1.454\pm0.023$ &  $D^+\to\pi^+\pi^0$ & $1.247\pm0.033$   & $0.973\pm0.017$ & $0.951\pm0.017$ \\
  $D^0\to \pi^0\pi^0$ & $0.826\pm0.025$ & $0.826\pm0.017$ & $0.826\pm0.017$  &  $D^+\to\pi^+\eta$ & $3.77\pm0.09$   & $3.30\pm0.09$  & $4.00\pm0.11$ \\
  $D^0\to \pi^0\eta$ & $0.63\pm0.06$ & $0.67\pm0.02$ & $0.91\pm0.02$ &  $D^+\to\pi^+\eta'$ & $4.97\pm0.19$   & $4.56\pm0.06$  & $4.68\pm0.05$\\
  $D^0\to \pi^0\eta'$ & $0.92\pm0.10$ & $1.00\pm0.02$ & $1.41\pm0.03$ &  $D^+\to K^+\ov K^0$ & $6.08\pm0.18$   & $8.44\pm0.18$  & $8.81\pm0.16$ \\
  $D^0\to \eta\eta$ & $2.11\pm0.19$ & $2.00\pm0.02$\footnotemark[1] & $1.81\pm0.02$\footnotemark[2] &  $D_s^+\to\pi^+ K^0$ & $2.18\pm0.10$   & $2.74\pm0.07$  & $2.57\pm0.06$ \\
  $D^0\to \eta\eta'$ & $1.01\pm0.19$ & $0.82\pm0.03$\footnotemark[3] & $0.77\pm0.03$\footnotemark[4] &  $D_s^+\to \pi^0K^+$ & $0.74\pm0.05$   & $0.54\pm0.02$  & $0.54\pm0.02$\\
  $D^0\to K^+K^-$ & $4.08\pm0.06$ & $4.08\pm0.04$ & $4.08\pm0.03$ &  $D_s^+\to K^+\eta$ & $1.73\pm0.06$   & $0.85\pm0.02$ & $0.84\pm0.02$ \\
  $D^0\to K^0\ov K^0$ & $0.282\pm0.010$ & $0.282\pm0.011$ & $0.282\pm0.013$ &  $D_s^+\to K^+\eta'$ & $2.64\pm0.24$   & $1.56\pm0.06$ & $1.67\pm0.07$ \\
\end{tabular}
\footnotetext[1]{The branching fraction becomes $2.17\pm0.03$ for the second solution of $W$-exchange.}
\footnotetext[2]{The branching fraction becomes $2.10\pm0.03$ for the second solution of $W$-exchange.}
\footnotetext[3]{The branching fraction becomes $1.79\pm0.07$ for the second solution of $W$-exchange.}
\footnotetext[4]{The branching fraction becomes $1.77\pm0.07$ for the second solution of $W$-exchange.}
\end{ruledtabular}}
\end{table}
%

We can fix the SU(3) breaking effects in the $W$-exchange amplitudes from the following four $D^0$ decay modes, $K^+K^-$, $\pi^+\pi^-$, $\pi^0\pi^0$ and $K^0\ov K^0$:
\be
\begin{split}
A(D^0\to \pi^+\pi^-)=\lambda_d (0.96T+E_d), &\quad&&  A(D^0\to \pi^0\pi^0)={1\over\sqrt{2}}\lambda_d (-0.78C+E_d),
\\
A(D^0\to K^+K^-)=\lambda_s (1.27T+E_s),  &\quad&& A(D^0\to K^0\ov K^0)=\lambda_d E_d+\lambda_s E_s,
\end{split}
\en
where $\lambda_q\equiv V^*_{cq}V_{uq}$ with $V_{qq'}$ denoting the $qq'$ element of the Cabibbo-Kobayashi-Maskawa matrix, and $E_q$ refers to the $W$-exchange amplitude associated with $c\bar u\to q\bar q$ ($q=d,s$).
A fit to the data  (see Table~\ref{tab:BRPP}) yields two possible solutions~\cite{Cheng:2019}
\be \label{eq:EdEs}
\begin{split}
{\rm I: }~~ &  E_d=1.244\, e^{i13.7^\circ}E ~, \qquad E_s=0.823\, e^{-i17.9^\circ}E
~; \\
{\rm II: }~~ &  E_d=1.244\, e^{i13.7^\circ}E ~, \qquad E_s=1.548\, e^{-i12.3^\circ}E
~,
\end{split}
\en
for $\phi=40.4^\circ$ and
\be \label{eq:EdEs}
\begin{split}
{\rm I: }~~ &  E_d=1.325\, e^{i12.9^\circ}E ~, \qquad E_s=0.795\, e^{-i17.7^\circ}E
~; \\
{\rm II: }~~ &  E_d=1.325\, e^{i12.9^\circ}E ~, \qquad E_s=1.665\, e^{-i13.5^\circ}E
~,
\end{split}
\en
for $\phi=43.5^\circ$.
In the $P\!P$ sector, we thus need SU(3) breaking in the $W$-exchange diagrams in order to induce the observed $D^0\to K^0_SK^0_S$ decay and explain the large rate difference between the $D^0\to K^+K^-$ and $D^0\to \pi^+\pi^-$ decays. Since the $W$-exchange and $W$-annihilation amplitudes are mainly governed by long-distance physics, their SU(3) breaking effects are obtained by fitting to the data, see Eq.~(\ref{eq:EdEs}).

Topological amplitudes for the SCS $D\to PP$ decays including perturbative SU(3) breaking effects in the $T$ and $C$ amplitudes and nonperturbative SU(3) breaking in the $E$ amplitude are summarized in Table~I of Ref.~\cite{Cheng:2019}. \footnote{In order to discuss \CP violation in charmed meson decays, we have included QCD penguin, penguin exchange and penguin annihilation in Table~I of Ref.~\cite{Cheng:2019} which can be neglected in the present study.}
The measured and fitted branching fractions are shown in Table~\ref{tab:BRPP} for the CF and SCS $D\to PP$ decays and in Table~\ref{tab:DCSPP} for the DCS decays.

\begin{table}[t]
\caption{Topological amplitude decompositions, experimental and predicted branching fractions for the DCS $D \to PP$ decays.  All branching fractions are quoted in units of $10^{-4}$. Here $\lambda_{ds}\equiv V^*_{cd}V_{us}$.
\label{tab:DCSPP}
}
\medskip
\footnotesize{
\begin{ruledtabular}
\begin{tabular}{l  c c c c  }
Mode & Amplitude & ${\cal B}_{\rm expt}$~\cite{PDG}  & ${\cal B}_{\rm theo}$ (PPa) & ${\cal B}_{\rm theo}$ (PPb)  \\
\hline
$D^0\to K^{+}\,\pi^-$          & $\lambda_{ds}(1.23 T + E)$ & $1.50\pm0.07$
&$1.74\pm0.02$   & $1.72\pm0.02$  \\
$D^0\to {K}^{0}\,\pi^0$             &$\frac{1}{\sqrt{2}}\lambda_{ds}(C - E)$ & --
&$0.66\pm0.01$       &$0.66\pm0.01$  \\           	
$D^0\to {K}^{0}\,\eta$               & $\lambda_{ds}\left[ {1\over \sqrt{2}}(C + E)\cos\phi - E\sin\phi\, \right]$ & --
&$0.273\pm0.001$    &$0.271\pm0.001$
\\
$D^0\to {K}^{0}\,\eta\,'$            &$\lambda_{ds}\left[ {1\over \sqrt{2}}(C + E)\sin\phi + E\cos\phi\, \right]$ & --
&$0.51\pm0.01$    &$0.50\pm0.01$  \\
$D^+\to K^{0}\,\pi^+$ & $\lambda_{ds}(C+0.71 A)$ & -- & $2.11\pm0.08$ & $2.18\pm0.07$ \\
$D^+\to K^{+}\,\pi^0$ & $\frac{1}{\sqrt{2}}\lambda_{ds}\left( 1.23 T-0.71A \right)$ & $2.08\pm0.21$ &  $2.54\pm0.06$ & $2.46\pm0.05$ \\
$D^+\to K^{+}\,\eta$ & $\lambda_{ds} \left[{1\over\sqrt{2}}(1.05 T+A)\cos\phi-0.81A\sin\phi\right]$  & $1.25\pm0.16$ & $1.04\pm0.01$ & $0.95\pm0.01$ \\
$D^+\to K^{+}\,\eta^{\prime}$ & $\lambda_{ds} \left[{1\over\sqrt{2}}(1.05T+A)\sin\phi+0.81A\cos\phi\right]$
& $1.85\pm0.20$ & $1.07\pm0.01$ & $1.39\pm0.07$ \\
$D_s^+\to K^{0}\,K^+$ & $\lambda_{ds}(1.27 T+1.03 C)$ & --
& $0.73 \pm 0.01$ & $0.71 \pm 0.01$  \\
\end{tabular}
\end{ruledtabular}}
\end{table}

\subsubsection{$K_S^0-K_L^0$ asymmetries}

Assuming that the double-primed amplitudes in the DCS sector are the same as that in the CF one, the calculated $D\to K_{S,L}^0P$ decays and their asymmetries defined by
\be \label{eq:R}
R(D,P)\equiv {\Gamma(D\to K_S^0 P)-\Gamma(D\to K_L^0 P)\over \Gamma(D\to K_S^0 P)+\Gamma(D\to K_L^0 P)}
\en
are summarized in Table~\ref{tab:PPKLKS}. It is expected that $D^0\to \ov K^0(\pi^0,\eta,\eta')$ and
$D^0\to  K^0(\pi^0,\eta,\eta')$ contribute constructively to $D^0\to K_S^0(\pi^0,\eta,\eta')$ and destructively to $D^0\to K_L^0(\pi^0,\eta,\eta')$ and hence,
\be
R(D^0,  (\pi^0,\eta,\eta'))=2\tan^2\theta_C=0.107\,.
\en
This prediction is in agreement with CLEO for $R(D^0,\pi^0)$~\cite{CLEO:D0KL} and with BESIII for $R(D^0, \eta^{(')})$~\cite{BESIII:D0KL}.

However, our prediction of $R(D^+,\pi^+)$ is opposite to experiment in sign and the calculated
$R(D_s^+, K^+)$ is too small compared to the data, though they are consistent if errors are taken into account. This can be traced back to the relative phase between $(C+A)$ and $(T+C)$ which is $94.5^\circ$. Consequently, $D^+\to \ov K^0\pi^+$ and $D^+\to K^0\pi^+$ will contribute destructively to $D^+\to K_S^0\pi^+$ and constructively to
$D^+\to K_L^0\pi^+$. This is opposite to the pattern observed experimentally. We find that if the phase difference is decreased slightly by $10^\circ$, that is, $(C''+A'')\to (C+A)e^{i 10^\circ}$ in $D^+\to K_{S,L}\pi^+$ and $(T''+C'')\to (T+C)e^{-i10^\circ}$ in $D_s^+\to K_{S,L}K^{+}$, then we will be able to accommodate both
$R(D^+,\pi^+)$ and $R(D_s^+, K^+)$.

\begin{table}[tp!]
\caption{Topological amplitude decompositions, branching fractions and $K_S^0-K_L^0$ asymmetries $R(D,P)$ for $D\to K_{S,L}^0P$ decays using Solution PPa denoted by 2023a in Table~\ref{tab:AmpPP}. Experimental results for $R$ are taken from Refs.~\cite{CLEO:D0KL,BESIII:D0KL,BESIII:DsKL}.
  \label{tab:PPKLKS}  }
\medskip
\scriptsize{
\begin{ruledtabular}
\begin{tabular}{l  c c c c  c  }
  Mode~~~ & Representation & ${\cal B}_{\rm expt}(\%)$~\cite{PDG} &  ${\cal B}_{\rm theo}(\%)$ &   $R_{\rm expt}$  &  $R_{\rm theo} $  \\
\hline
  $D^0\to K_S^0\pi^0$ &  ${1\over 2}(\lambda_{sd}-\lambda_{ds})(C-E)$ &  $1.240\pm0.022$  &  $1.282\pm0.017$  &  $0.108\pm0.035$ & $0.107\pm0.009$  \\
  $D^0\to K_L^0\pi^0$ &  ${1\over 2}(\lambda_{sd}+\lambda_{ds})(C-E)$ &  $0.976\pm0.032$  &  $1.035\pm0.014$   \\
  $D^0\to K_S^0\eta$ & ${1\over 2}(\lambda_{sd}-\lambda_{ds})[(C+E)\cos\phi-{1\over\sqrt{2}}E\sin\phi]$ & $0.509\pm0.013$ &   $0.531\pm0.006$   & $0.080\pm0.022$ & $0.107\pm0.008$  \\
  $D^0\to K_L^0\eta$ & ${1\over 2}(\lambda_{sd}+\lambda_{ds})[(C+E)\sin\phi-{1\over\sqrt{2}}E\cos\phi]$ & $0.434\pm0.016$ & $0.429\pm0.005$  \\
  $D^0\to K_S^0\eta'$ & ${1\over 2}(\lambda_{sd}-\lambda_{ds})[(C+E)\sin\phi+{1\over\sqrt{2}}E\cos\phi]$ & $0.949\pm0.032$ &    $0.983\pm0.024$    & $0.080\pm0.023$  & $0.107\pm0.017$ \\
  $D^0\to K_L^0\eta'$ & ${1\over 2}(\lambda_{sd}+\lambda_{ds})[(C+E)\sin\phi+{1\over\sqrt{2}}E\cos\phi]$ & $0.812\pm0.035$ &  $0.794\pm0.019$ \\
  $D^+\to K_S^0\pi^+$ &  ${1\over \sqrt{2}}[\lambda_{sd}(T+C)-\lambda_{ds}(C+A])]$ &  $1.562\pm0.031$   & $1.524\pm0.030$  & $0.022\pm0.024$ & $-0.013\pm0.013$ \\
   $D^+\to K_L^0\pi^+$ &  ${1\over\sqrt{ 2}}[\lambda_{sd}(T+C)+\lambda_{ds}(C+A)]$ &  $1.460\pm0.053$  & $1.563\pm0.029$  \\
  $D_s^+\to K_S^0K^+$ &  ${1\over \sqrt{2}}[\lambda_{sd}(C+A)-\lambda_{ds}(T+C)]$ &  $1.450\pm0.035$  &  $1.462\pm0.044$ & $-0.021\pm0.025$ & $-0.006\pm0.020$  \\
   $D_s^+\to K_L^0K^+$ &  ${1\over \sqrt{2}}[\lambda_{sd}(C+A)+\lambda_{ds}(T+C)]$ &  $1.485\pm0.060$ & $1.478\pm0.040$  \\
\end{tabular}
\end{ruledtabular}}
\end{table}
In  the so-called
factorization-assisted topological (FAT) approach~\cite{Li:2012},  $R(D^+,\pi^+)$ and $R(D_s^+, K^+)$ are predicted to be $0.025\pm0.008$ and $0.012\pm0.006$, respectively~\cite{Wang}. While the former agrees with experiment, the latter is wrong in sign.

\section{$D\to V\!P$ decays \label{sec:DtoVP}}

\subsubsection{Cabibbo-favored $D\to V\!P$ decays}

For $D\to V\!P$ decays, there exist two different sets of topological diagrams since the spectator quark of the charmed meson may end up in the pseudoscalar or vector meson.  A subscript of $P$ or $V$ is attached to the flavor amplitudes and the associated strong phases to denote respectively whether the spectator quark in the charmed meson ends up in the pseudoscalar or vector meson in the final state.

As mentioned in passing, for the CF decay modes involving a neutral kaon $K_S^0$ or $K_L^0$, it was customary to use the relation $\Gamma(\ov K^0)=2\Gamma(K_S^0)$, but this relation can be invalidated by the interference between CF and DCS amplitudes. Just as the $P\!P$ decays,
we also prefer to apply the relation $\B(D\to \ov K^0 V)\cong \B(D\to K_S^0 V)+\B(D\to K_L^0 V)$ in the $V\!P$ sector.   Unfortunately, we have the data of $\B(D\to K_L^0 V)$  for $V=\omega$ and $\phi$, but not for $V=\rho^0, \rho^+$ and $K^{*+}$. Since the double-primed topological amplitudes in DCS decays are not the same as unprimed ones extracted from CF modes, as we shall discuss later, we will fit to the measured rates of $D^0\to \ov K^0(\omega,\phi)$, $D^+\to K_S^0\rho^+$ and $D_s^+\to K_S K^{*+}$, but not to $D^0\to K_S^0\rho^0$. Owing to the absence of the data on $\B(D^0\to K_L^0\rho^0)$, we shall assume that the experimental branching fraction of $D^0\to \ov K^0\rho^0$ be two times $\B(D^0\to K_S^0\rho^0)$.

\begin{table}[tp!]
\caption{Flavor amplitude decompositions, experimental branching fractions, and predicted branching fractions for CF $D \to V\!P$ decays.  Data are taken from the Particle Data Group~\cite{PDG} unless specified otherwise.  The column of ${\cal B}_{\rm theo}$ shows predictions based on solution (F4) presented in Table~\ref{tab:CFVPsolution_1} and solution (F1') in Table~\ref{tab:CFVPsolution_2}.  All branching fractions are quoted in units of \%.
}
\vspace{6pt}
\footnotesize{
\begin{ruledtabular}
\begin{tabular}{l l l c c c c c c}
Meson & Mode & Amplitude decomposition
& ${\cal B}_{\rm expt}$ & ${\cal B}_{\rm theo}$ (F4) & ${\cal B}_{\rm theo}$ (F1')
\\
\hline
$D^0$  & $K^{*-}\,\pi^+$          & $\lambda_{sd}(T_V + E_P)$
&$5.34\pm0.41$                    &$5.45\pm0.34$	 &$5.37\pm0.33$	
\\
&$K^-\,\rho^+$                      & $\lambda_{sd}(T_P + E_V)$
&$11.2 \pm 0.7$                    &$11.4\pm0.8$	 &$11.3\pm0.7$	
\\
& $\ol{K}^{*0}\,\pi^0$             &$\frac{1}{\sqrt{2}}\lambda_{sd}(C_P - E_P)$
&$3.74\pm0.27$                     &$3.61\pm0.18$	 &$3.70\pm0.18$
\\
& $\ov {K}^0\,\rho^0$              & $\frac{1}{\sqrt{2}}\lambda_{sd}(C_V -
                                    E_V)$
&$1.26^{+0.12}_{-0.16}$           &$1.25\pm0.09$	 &$1.25\pm0.09$	
\\
& $\ol{K}^{*0}\,\eta$               & $\lambda_{sd}\left[ {1\over \sqrt{2}}(C_P + E_P)\cos\phi - E_V\sin\phi\, \right]$
&$1.41\pm0.12$                    &$1.35\pm0.06$	  &$1.41\pm0.07$		
\\
& $\ol{K}^{*0}\,\eta\,'$            &$\lambda_{sd}\left[ {1\over \sqrt{2}}(C_P + E_P)\sin\phi + E_V\cos\phi\, \right]$
&$<0.10$                            &$0.0055\pm0.0004$   &$0.0043\pm0.0003$
\\
& $\ol{K}^0\,\omega$             & $\frac{1}{\sqrt{2}}\lambda_{sd}(C_V + E_V)$
&$2.22 \pm 0.12$                 &$2.29\pm0.11$	  &$2.29\pm0.11$	
\\
& $\ol{K}^0\,\phi$                   &$\lambda_{sd}E_P$
&$0.825\pm0.061$  			& $0.830\pm0.034$		& $0.828\pm0.034$	
\\
\hline
$D^+$
& $\ol{K}^{*0}\,\pi^+$  & $\lambda_{sd}(T_V + C_P)$
&$1.57\pm0.13$                   &$1.58\pm0.13$	   &$1.58\pm0.13$		
\\
&${K}_S^0\,\rho^+$               & ${1\over\sqrt{2}}[\lambda_{sd}(T_P + C_V)-\lambda_{ds}(C_V+A_P)]$
&$6.14^{+0.60}_{-0.35}$             &$6.38\pm0.44$	  &$6.26\pm0.52$		
\\
\hline
$D_s^+$
& $\ol{K}^{*0}\,K^+$& $\lambda_{sd}(C_P + A_V)$
&$3.79\pm0.09$                    &$3.80\pm0.10$	 &$3.79\pm0.09$
\\
&${K}_S^0\,K^{*+}$              & ${1\over\sqrt{2}}[\lambda_{sd}(C_V + A_P)-\lambda_{ds}(T_P+C_V)]$
&$0.77 \pm 0.07$\footnotemark[1]                   &$0.79\pm0.04$	 &$0.78\pm0.03$		
\\
&$\rho^+\,\pi^0$                   & $\frac{1}{\sqrt{2}}\lambda_{sd}(A_P - A_V)$
&---                                  &$0.012\pm0.003$	  &$0.011\pm0.002$	
\\
&$\rho^+\,\eta$                    & $\lambda_{sd}\left[ {1\over\sqrt{2}}( A_P + A_V)\cos\phi-T_P \sin\phi \right]$
&$8.9 \pm 0.8$                     &$9.25\pm0.35$   &$8.75\pm0.31$
\\
&$\rho^+\,\eta\,'$                 & $\lambda_{sd} \left[{1\over\sqrt{2}}( A_P + A_V)\sin\phi+T_P\cos\phi \right]$
&$5.8\pm1.5$   						&$3.24\pm0.11$	&$3.60\pm0.11$	
\\
& $\pi^+\,\rho^0$                 & $\frac{1}{\sqrt{2}}\lambda_{sd}(A_V - A_P)$
 &$0.0112\pm0.0013$\footnotemark[2]                &$0.011\pm0.003$	   &$0.011\pm0.002$	
\\
& $\pi^+\,\omega$                &$\frac{1}{\sqrt{2}}\lambda_{sd}(A_V + A_P)$
&$0.238\pm 0.015$\footnotemark[3]                  &$0.24\pm0.01$	 &$0.24\pm0.01$
\\
& $\pi^+\,\phi$                      &$\lambda_{sd}T_V$
&$4.50 \pm 0.12$                      &$4.49\pm0.11$	  &$4.50\pm0.11$	
\\
\end{tabular}
\label{tab:VPCF}
\footnotetext[1]{This is the average of the branching fractions $(2.7\pm0.6)\%$~\cite{CLEO:1989}, $(0.612\pm 0.099)\%$~\cite{BESIII:DsKsKppi} and $(0.927\pm 0.099)\%$~\cite{BESIII:DsKsKspi}.}
\footnotetext[2]{This is from the new LHCb analysis of $D_s^+\to \pi^+\pi^+\pi^-$ decays~\cite{LHCb:Dstopipipi}.}
\footnotetext[3]{The new LHCb measurement of $D_s^+\to \pi^+\omega$~\cite{LHCb:Dstopipipi} is taken into account in the world average.}
\end{ruledtabular}}
\end{table}

The partial decay width of the $D\to V\!P$ decay can be expressed in two different ways:
\begin{equation}
\Gamma(D\to V\!P)=\frac{p_c}{8\pi m_D^2}\sum_{\rm pol}|{{\cal  M}}|^2,
\label{decaywidthC}
\end{equation}
or
\begin{equation}
\Gamma(D\to V\!P)=\frac{p_c^3}{8\pi m_V^2}|\tilde{{\cal M}}|^2 ~,
\label{decaywidthA}
\end{equation}
with ${\cal M}=\tilde{\cal M}(\epsilon\cdot p_D)$. Because the additional SU(3)-breaking factor in phase space has been taken care of, we prefer to use Eq.~\eqref{decaywidthA}.
By performing a $\chi^2$ fit to the CF $D\to V\!P$ decay rates (see Table~\ref{tab:VPCF}), we extract the magnitudes and strong phases of the topological amplitudes $T_V,C_V,E_V,A_V$ and $T_P,C_P,E_P,A_P$ from the measured partial widths through Eq.~(\ref{decaywidthA}) and find many possible solutions with local $\chi^2$ minima.  Here we take the convention that all strong phases are defined relative to the $T_V$ amplitude. We obtain five best $\chi^2$-fit solutions (F1)-(F5) in
Table~\ref{tab:CFVPsolution_1} for $\phi=40.4^\circ$ and (F1')-(F5') in Table~\ref{tab:CFVPsolution_2} for $\phi=43.5^\circ$, where we have restricted ourselves to those with $\chi^2_{\rm min}<10$. The topological amplitudes of all these solutions respect the hierarchy pattern:
\be \label{eq:tree_hierarchy}
|T_P|>|T_V|>|C_{P}|>|E_P|>|C_{V}|\gsim|E_V|>|A_{P,V}|.
\en
This is slightly different from the hierarchy pattern previously found by us in 2021~\cite{Cheng:2021}:
\be
|T_P|>|T_V|\gsim|C_{P}|>|C_{V}|\gsim|E_P|>|E_V|>|A_{P,V}|.
\en

\begin{table}[t]
\caption{Solutions for the topological amplitudes in CF $D\to V\!P$ decays with $\chi^2_{\rm min} \le 10$ obtained using Eq.~(\ref{decaywidthA}) and $\phi=40.4^\circ$.  The amplitude sizes are quoted in units of  $10^{-6}(\epsilon\cdot p_D)$ and the strong phases in units of degrees.
}
\label{tab:CFVPsolution_1}
\medskip
\footnotesize{
\begin{ruledtabular}
\begin{tabular}{c c c c c c c c c c c}
Set       &$|T_V|$                   &$|T_P|$            &$\delta_{T_P}$          &$|C_V|$                          &$\delta_{C_V}$          &$|C_P|$                  &$\delta_{C_P}$                &$|E_V|$                         &$\delta_{E_V}$     \\
          &$|E_P|$                            &$\delta_{E_P}$   &$|A_V|$                           &$\delta_{A_V}$             &$|A_P|$ &$\delta_{A_P}$ &$\chi^2_{\rm min}$ &fit quality\\
\hline
(F1)  &$2.17\pm0.03$ & $3.56\pm0.06$ & $301^{+5}_{-4}$ & $1.47\pm0.03$ & $134\pm2$ & $2.07\pm0.02$ &$201\pm1$ & $1.02\pm0.04$ & $63\pm2$  \\
            &$1.65\pm0.04$ & $107\pm2$ & $0.21\pm0.01$ & $320\pm2$ & $0.24\pm0.01$  & $351\pm2$            & 2.51 & 47.31\%  \\ 
\hline
(F2)  &$2.17\pm0.03$ & $3.59\pm0.06$ &  $32^{+4}_{-5}$ & $1.52\pm0.03$ & $197\pm2$ & $1.99\pm0.02$ & $201\pm1$ & $0.93\pm0.04$ & $267\pm2$  \\
            & $1.65\pm0.03$ & $106\pm3$ & $0.18\pm0.01$ & $337^{+6}_{-3}$ & $0.26\pm0.01$  & $345^{+4}_{-7}$           & 3.10 & 37.61\% \\ 
\hline
(F3)  &$2.17\pm0.03$ & $3.50\pm0.06$ &$183\pm4$ & $1.39\pm0.03$ &$13\pm2$ & $1.97\pm0.02$ &$159\pm1$    & $1.12\pm0.04$ & $301\pm2$  \\
            &$1.65\pm0.04$ & $254\pm2$ &$0.23\pm0.01$ &$279\pm1$ & $0.22\pm0.01$  &$255^{+2}_{-1}$           & 3.29 & 37.86\% \\ 
\hline
(F4)  &$2.17\pm0.03$ & $3.51\pm0.06$ &$14\pm4$ &$1.38\pm0.03$ &$185\pm2$ &$2.00\pm0.02$ &$201\pm1$ &$1.14\pm0.04$ &$257\pm2$  \\
            &$1.66\pm0.03$ &$107\pm3$ &$0.26\pm0.01$ &$78\pm3$ &$0.18\pm0.01$  &$67^{+4}_{-3}$           & 3.87& 27.60\% \\ 
\hline
(F5)  &$2.17\pm0.03$ &$3.29\pm0.06$ &$178\pm4$ &$1.17\pm0.03$ &$357\pm2$ &$1.99\pm0.02$ &$201\pm1$ &$1.36\pm0.03$ &$69\pm2$  \\
            &$1.66\pm0.03$ &$107\pm2$ &$0.24\pm0.01$ &$324^{+1}_{-2}$ &$0.21\pm0.01$  &$301^{+2}_{-1}$           & 5.20 & 15.75\% \\ 
\end{tabular}
\end{ruledtabular}
\label{fitting}}
\end{table}

\begin{table}[t!]
\caption{Same as Table~\ref{tab:CFVPsolution_1} except for $\phi=43.5^\circ$.
}
\label{tab:CFVPsolution_2}
\medskip
\footnotesize{
\begin{ruledtabular}
\begin{tabular}{c c c c c c c c c c c}
Set       &$|T_V|$                   &$|T_P|$            &$\delta_{T_P}$          &$|C_V|$                          &$\delta_{C_V}$          &$|C_P|$                  &$\delta_{C_P}$                &$|E_V|$                         &$\delta_{E_V}$     \\
          &$|E_P|$                            &$\delta_{E_P}$   &$|A_V|$                           &$\delta_{A_V}$             &$|A_P|$ &$\delta_{A_P}$ &$\chi^2_{\rm min}$ &fit quality\\
\hline
(F1')  &$2.17\pm0.03$ &$3.58\pm0.06$ &$327^{+5}_{-4}$ &$1.53\pm0.03$ &$161\pm2$ &$2.06\pm0.02$ &$159\pm1$ &$0.92\pm0.04$ &$92\pm2$  \\
            &$1.65\pm0.03$ &$253\pm3$ &$0.20\pm0.01$ &$329\pm2$ &$0.25\pm0.01$  &$349\pm2$            & 2.24 & 52.44\%  \\ 
\hline
(F2')  &$2.17\pm0.03$ &$3.55\pm0.06$ &$65^{+4}_{-5}$ &$1.49\pm0.03$ &$232\pm2$ &$2.05\pm0.02$ &$159\pm1$ &$1.00\pm0.04$ &$303\pm2$  \\
            &$1.65\pm0.04$ &$253\pm2$ &$0.25\pm0.01$ &$19\pm2$ &$0.20\pm0.01$  &$41\pm2$           & 2.49 & 47.75\% \\ 
\hline
(F3')  &$2.17\pm0.03$ &$3.50\pm0.06$ &$143^{+5}_{-4}$ &$1.41\pm0.03$ &$333\pm2$ &$2.00\pm0.02$ &$201\pm1$    &$1.10\pm0.04$ &$261\pm2$  \\
            &$1.66\pm0.03$ &$106\pm3$ &$0.24\pm0.01$ &$77\pm1$ &$0.21\pm0.01$  &$100^{+1}_{-2}$           & 3.28 & 35.06\% \\ 
\hline
(F4')  &$2.17\pm0.03$ &$3.54\pm0.06$ &$165^{+4}_{-5}$ &$1.49\pm0.03$ &$333\pm2$ &$1.91\pm0.02$ &$201\pm1$ &$0.99\pm0.04$ &$43\pm2$  \\
            &$1.66\pm0.04$ &$105\pm2$ &$0.22\pm0.01$ &$99\pm1$ &$0.23\pm0.01$  &$123^{+1}_{-2}$       & 4.34 & 22.66\% \\ 
\hline
(F5')  &$2.17\pm0.03$ &$3.25\pm0.06$ &$173\pm4$ &$1.15\pm0.03$ &$353\pm2$ &$1.92\pm0.02$ &$201\pm1$ &$1.37\pm0.03$ &$65\pm2$  \\
            &$1.66\pm0.04$ &$106\pm2$ &$0.20\pm0.01$ &$310\pm2$ &$0.25\pm0.01$  &$288\pm2$           & 7.94 & 4.72\% \\ 
\end{tabular}
\end{ruledtabular}
\label{fitting}}
\end{table}


Comparing Tables \ref{tab:CFVPsolution_1} and \ref{tab:CFVPsolution_2} with the previous five best solutions (S1')-(S5') obtained in Table II of~\cite{Cheng:2021}, it is evident that (i) the magnitudes of $|C_V|$ and $|A_V-A_P|$ are decreased, whereas $|E_V|$ is increased, and (ii) the uncertainties in the magnitudes and phases of $A_{P,V}$ are significantly improved. These can be traced back to the improved branching fractions of CF $D\to V\!P$ modes shown in Table~\ref{tab:VPCF}. The branching fraction $\B(D_s^+\to \ov K^0 K^{*+})=(2.7\pm0.6)\%$ reported by CLEO in 1989~\cite{CLEO:1989} has been found to be much smaller, $(0.77\pm0.07)\%$, by BESIII~\cite{BESIII:DsKsKspi,BESIII:DsKsKppi} very recently with a greater precision. Since the annihilation amplitude $A_P$ is very suppressed compared to $C_V$, this implies that the magnitude of $C_V~(F4)$, $1.38\pm0.03$, (see Table~\ref{tab:CFVPsolution_1}) is smaller than $|C_V(S3')|=1.69\pm0.04$~\cite{Cheng:2021}. Since $A(D^0\to\ov K^0\rho^0)\propto (C_V-E_V)$, a decrease in $|C_V|$ implies an increase in $|E_V|$. A new measurement of $D_s^+\to\pi^+\rho^0$ indicates that its branching fraction is significantly improved from $(1.2\pm0.6)\times 10^{-4}$~\cite{PDG} to $(1.12\pm0.13)\times 10^{-4}$~\cite{LHCb:Dstopipipi} with a much better precision. This in turn implies a smaller $|A_V-A_P|$ and more precise values of $A_V$ and $A_P$. The extremely small branching fraction of $D_s^+\to\pi^+\rho^0$ compared to that of $D_s^+\to \pi^+\omega$ (see Table~\ref{tab:VPCF}) implies that $A_V$ and $A_P$ should be comparable in magnitude and roughly parallel to each other with a phase difference not more than $25^\circ$.

\subsubsection{Singly-Cabibbo-suppressed $D\to V\!P$ decays}

Although the five solutions generally fit the CF modes well (see Table~\ref{tab:VPCF}), there is one exception, namely, $D_s^+\to\rho^+\eta'$, whose prediction is smaller than the experimental result. As explained in Ref.~\cite{Cheng:2016}, this mode has a decay amplitude respecting a sum rule:
\be
A(D_s^+\to \pi^+\omega)=\cos\phi\, A(D_s^+\to\rho^+\eta)+\sin\phi\, A(D_s^+\to\rho^+\eta').
\en
The current data of $\B(D_s^+\to\pi^+\omega)$ and $\B(D_s^+\to\rho^+\eta)$ give the bounds $1.6\% < \B(D_s^+\to\rho^+\eta') < 3.9\%$ at $1\sigma$ level, significantly lower than the current central value. A recent update by BESIII  yields $\B(D_s^+\to\rho^+\eta')=(6.15\pm0.31)\%$~\cite{BESIII:Dsrhoetap}. This seems to imply that it is necessary to take into account the extra contribution from the flavor-singlet topological amplitude $S$ available for the $\eta'$ meson~\cite{Cheng:2011qh}.

The five solutions (F1)-(F5) or (F1')-(F5'), although describing the CF decays well, may lead to very different predictions for some of SCS modes. In particular, the decays $D^{0,+}\to\pi^{0.+}\omega$, $D^{0,+}\to \pi^{0,+}\rho^0$ and $D^0\to\eta\omega$ are very useful in discriminating among the different solutions. Their topological amplitudes read
\be \label{eq:D+pirho}
\begin{split}
A(D^+\to \pi^+\rho^0) &= {1\over \sqrt{2}}\lambda_d (T_V+C_P-A_P+A_V),
\\
A(D^+\to \pi^+\omega) &= {1\over \sqrt{2}}\lambda_d (T_V+C_P+A_P+A_V),
\end{split}
\en
and
\be \label{eq:D0pirho}
\begin{split}
A(D^0\to \pi^0\omega) &= {1\over 2}\lambda_d(C_V-C_P+E_P+E_V),
\\
A(D^0\to \pi^0\rho^0) &= {1\over 2}\lambda_d(C_V+C_P-E_P-E_V),
\\
A(D^0\to \eta\omega) &= {1\over 2}\lambda_d(C_V+C_P+E_P+E_V)\cos\phi-{1\over\sqrt{2}}\lambda_s C_V\sin\phi.
\end{split}
\en
Experimental measurements indicate that (in units of $10^{-3}$, see Table~\ref{tab:SCSVP})
\be \label{eq:D+pirho_b}
\begin{split}
 \B(D^+\to\pi^+\rho^0)=0.83\pm0.14&>\B(D^+\to\pi^+\omega)=0.28\pm0.06\,,
 \\
 \B(D^0\to\pi^0\rho^0)=3.86\pm0.23&> \B(D^0\to\eta\omega)=1.98\pm0.18\,,
 \\
  &\gg \B(D^0\to\pi^0\omega)=0.117\pm0.035\,.
\end{split}
\en
Since $A_V$ and $A_P$ are comparable in magnitude and roughly parallel to each other, it is tempting to argue from Eq.~(\ref{eq:D+pirho}) that $D^+\to\pi^+\omega$ should have a rate larger than $D^+\to\pi^+\rho^0$, which is opposite to the experimental finding. Since $C_P$ is comparable to $T_V$ in magnitude, there is a large cancellation between $T_V$ and the real part of $C_P$. In general, the experimental constraint from Eq.~(\ref{eq:D+pirho_b}) can be satisfied provided that the imaginary part of $C_P$ has a sign opposite to the imaginary part of $A_V$ or $A_P$. We find explicitly that the only allowed solutions are (F3), (F4), (F3') and (F5').

\begin{table}[tp!]
\caption{Branching fractions (in units of $10^{-3}$) of SCS $D\to V\!P$ decays. The predictions have taken into account SU(3) breaking effects under solution (i) in Table~\ref{tab:EPEV}.
  \label{tab:SCSVP}  }
\medskip
\scriptsize{
\begin{ruledtabular}
\begin{tabular}{l  c c c l c c c}
  Mode~~~ & ${\cal B}_{\rm exp}$~~ &  ${\cal B}_{\rm theo}$ (F4) &  ${\cal B}_{\rm theo}$ (F1')
  & Mode~~~ & ${\cal B}_{\rm exp}$~~ &  ${\cal B}_{\rm theo}$ (F4) &  ${\cal B}_{\rm theo}$ (F1')
\\
\hline
  $D^0\to\pi^+ \rho^-$ & $5.15\pm0.25$ & $5.42 \pm 0.12$ & $5.23 \pm 0.18$ &  $D^0\to\pi^0 \omega$  & $0.117\pm0.035$ & $0.157\pm0.015$  & $0.153\pm0.021$ \\
  $D^0\to\pi^- \rho^+$ & $10.1\pm0.4$  & $10.6 \pm 0.5$  & $10.2 \pm 0.6$ & $D^0\to\pi^0 \phi$ &  $1.17 \pm 0.04$ & $0.93\pm0.02$  & $0.99\pm0.02$ \\
  $D^0\to \pi^0 \rho^0$ & $3.86\pm0.23$  & $2.86 \pm 0.06$ & $3.38 \pm 0.10$ & $D^0\to\eta \omega$  & $1.98\pm0.18$ & $1.71\pm0.05$ & $1.99\pm0.06$  \\
  $D^0\to K^+ K^{*-}$ & $1.65\pm0.11$  & $1.65 \pm 0.04$ & $1.55 \pm 0.04$ & $D^0\to\eta\,' \omega$ &--- & $0.017\pm0.001$ & $0.009\pm0.001$ \\
  $D^0\to K^- K^{*+}$ & $4.56\pm0.21$  & $4.57 \pm 0.22$ & $4.56 \pm 0.15$ & $D^0\to\eta \phi$ &  $0.181\pm0.046$ & $0.175\pm0.007$ & $0.186\pm0.004$ \\
  $D^0\to K^0 \ol{K}^{*0}$ & $0.246\pm0.048$  & $0.246\pm0.011$  & $0.246\pm0.021$ & $D^0\to\eta \rho^0$ & ---  & $0.26\pm0.02$  & $0.25\pm0.02$  \\
  $D^0\to\ol{K}^0 K^{*0}$ & $0.336\pm0.063$  & $0.336\pm0.021$ & $0.336\pm0.015$ & $D^0\to\eta\,' \rho^0$ & ---  & $0.059\pm0.002$ & $0.059\pm0.002$ \\
\hline
  $D^+\to\pi^+ \rho^0$ & $0.83\pm0.14$ & $0.55 \pm 0.06$ & $0.57 \pm 0.05$ & $D^+\to\eta \rho^+$ & --- & $0.38\pm0.18$ & $0.36\pm0.19$ \\
  $D^+\to\pi^0 \rho^+$ & --- & $5.20\pm0.33$  & $5.25\pm0.38$ &  $D^+\to\eta\,' \rho^+$ & --- & $0.97\pm0.03$  & $1.12\pm0.03$ \\
  $D^+\to\pi^+ \omega$ & $0.28\pm0.06$ & $0.31\pm0.05$ & $0.88\pm0.07$ & $D^+\to K^+ \ol{K}^{*0}$  & $3.71\pm0.18$ & $5.78\pm0.15$ & $5.26\pm0.14$ \\
  $D^+\to\pi^+ \phi$ & $5.70\pm0.14$  & $4.74\pm0.10$ & $5.03\pm0.10$ & $D^+\to\ol{K}^0 K^{*+}$  & $17.3 \pm 1.8$  & $15.8\pm0.5$  & $15.6\pm0.5$  \\
\hline
  $D_s^+\to\pi^+ K^{*0}$ & $2.55\pm0.35$\footnotemark[1] & $2.06\pm0.06$ & $1.58\pm0.05$ & $D_s^+\to\eta K^{*+}$ & --- & $0.37\pm0.07$ & $0.39\pm0.09$ \\
  $D_s^+\to\pi^0 K^{*+}$ &  $0.75\pm0.25$\footnotemark[2] & $0.71\pm0.03$ & $0.67\pm0.03$ & $D_s^+\to\eta\,' K^{*+}$  & ---  & $0.40\pm0.02$ & $0.42\pm0.02$ \\
  $D_s^+\to K^+ \rho^0$ & $2.17\pm0.25$  & $1.01 \pm 0.03$ & $1.11 \pm 0.03$ &  $D_s^+\to K^+ \omega$ & $0.99\pm0.15$ & $1.16\pm0.03$ & $1.17\pm0.03$ \\
  $D_s^+\to K^0 \rho^+$ & $5.46\pm0.95$\footnotemark[2] & $7.54\pm0.27$  & $7.30\pm0.26$& $D_s^+\to K^+ \phi$  & $0.18\pm0.04$ & $0.11\pm0.01$ & $0.29\pm0.02$ \\
\end{tabular}
\footnotetext[1]{The new measurement of $\B(D_s^+\to \pi^+K^{*0})=(2.71\pm0.72\pm0.30)\times 10^{-3}$ from BESIII~\cite{BESIII:DsKSpippi0} is taken into account in the world average.}
\footnotetext[2]{Data from BESIII~\cite{BESIII:DsKSpippi0}, but not cited in PDG~\cite{PDG}.}
\end{ruledtabular}}
\end{table}
%

As for the $\pi^0\rho^0,\pi^0\omega$ and $\eta\omega$ modes, we see from Eq. (\ref{eq:D0pirho}) that the smallness of $\B(D^0\to\pi^0\omega)$, the sizable $\B(D^0\to\eta\omega)$ and the large $\B(D^0\to\pi^0\rho^0)$ imply that the strong phases of $C_V$ and $C_P$ should be close to each other~\cite{Cheng:2019}. An inspection of Tables \ref{tab:CFVPsolution_1} and \ref{tab:CFVPsolution_2} shows that the phase difference between $C_V$ and $C_P$ is small only for solutions (F2), (F4) and (F1'). All the other solutions yield $\B(D^0\to\pi^0\omega)>\B(D^0\to\eta\omega)$, in contradiction to the observation of $\B(D^0\to\pi^0\omega)\ll\B(D^0\to\eta\omega)$.  However, as just noted in passing, (F2) and (F1') will lead to the prediction of $\B(D^+\to\pi^+\omega)/\B(D^+\to\pi^+\rho^0)>1$, not consistent with the experiment (see Table~\ref{tab:SCSVP}).

\begin{table}[t]
\caption{The parameters $e_{V,P}^{d,s}$ and the phases $\delta e_{V,P}^{d,s}$ (in units of degrees) describing SU(3) breaking effects in the $W$-exchange amplitudes $E_V^{d,s}$ and $E_P^{d,s}$ for solutions (F4) (upper rows) and (F1') (lower rows), respectively, with the corresponding values of $\chi^2$ being 22.3 and 5.6 per degree of freedom.
  \label{tab:EPEV}}
  \medskip
\begin{ruledtabular}
\begin{tabular}{ l c  c c c c c c c}
 &  $e_V^d$ & $\delta e_V^d$ & $e_P^d$ & $\delta e_P^d$ & $e_V^s$ & $\delta e_V^s$ & $e_P^s$ & $\delta e_P^s$  \\
\hline
(i) & 0.53 & 18 & 0.21 & 282 & 0.78 & 30 & 0.16 & 340 \\
(ii) & 0.53 & 18 & 0.21 & 282 & 0.78 & 30 & 0.84 & 146 \\
(iii)& 0.53 & 18 & 0.21 & 282 & 0.92 & 207 & 0.16 & 340 \\
(iv) & 0.53 & 18 & 0.21 & 282 & 0.92 & 207 & 0.84 & 146 \\
\hline
(i) & 0.92 & 16 & 0.50 & 335 & 0.30 & 50 & 0.02 & 294 \\
(ii) & 0.92 & 16  & 0.50 & 335 & 0.30 & 50 & 0.87 & 145 \\
(iii)& 0.92 & 16  & 0.50 & 335 & 1.47 & 223 & 0.02 & 294 \\
(iv) & 0.92 & 16  & 0.50 & 335 & 1.47 & 223 & 0.87 & 145 \\
\end{tabular}
\end{ruledtabular}
\end{table}

Just like the $P\!P$ sector, we also need SU(3) breaking in the $W$-exchange amplitudes in the $V\!P$ sector because the ratios, $\Gamma(D^0\to K^+K^{*-})/\Gamma(D^0\to \pi^+\rho^-)=0.32\pm0.03$ and $\Gamma(D^0\to K^-K^{*+})/\Gamma(D^0\to \pi^-\rho^+)=0.45\pm0.03$~\cite{PDG}, deviate sizably from unity expected in the SU(3) limit. Also, the predicted rates of $D^0\to K^0\ol K^{*0}$ and $D^0\to \ol K^0 K^{*0}$ modes are too large by one order of magnitude compared to the experiment~\cite{Cheng:2019}. Writing
\be \label{eq:SU(3)inW}
\begin{split}
E_V^d=e_V^d e^{i\delta e_V^d}E_V, \quad E_V^s=e_V^s e^{i\delta e_V^s}E_V,
\\
E_P^d=e_P^d e^{i\delta e_P^d}E_P, \quad E_P^s=e_P^se^{i\delta e_P^s} E_P,
\end{split}
\en
and replacing $\lambda_d E_{V,P}$ by $\lambda_d E_{V,P}^d$ and $\lambda_s E_{V,P}$ by $\lambda_s E_{V,P}^s$ in the SCS $D^0\to V\!P$ decay amplitudes (see Ref.~\cite{Cheng:2019} for detail), we fit the eight unknown parameters $e_V^d,e_P^d,e_V^s,e_P^s$ and $\delta e_V^d,\delta e_P^d,\delta e_V^s,\delta e_P^s$ using the branching fractions of the following eight modes: $D^0\to \pi^+\rho^-,\pi^-\rho^+,\pi^0\rho^0,\pi^0\omega$ and $D^0\to K^+K^{*-}.K^-K^{*+},K^0\ol K^{*0},\ol K^0K^{*0}$. Table~\ref{tab:EPEV} shows the four solutions of SU(3) breaking effects in the $W$-exchange amplitudes for solutions (F4) and (F1').  In the SU(3) limit, $e_{V,P}^{d,s}=1$ and $\delta e_{V,P}^{d,s}=0$. Unlike solution (S3') found previously in Ref.~\cite{Cheng:2021} which leads to exact solutions for $e_{V,P}^{d,s}$ and $\delta e_{V,P}^{d,s}$ (i.e., $\chi^2=0$), here we do not have exact solutions and the values of $\chi^2$ are 22.3 and 5.6 per degree of freedom for solutions (F4) and (F1'), respectively. Indeed, we see from Table~\ref{tab:SCSVP} that the measured branching fraction of $D^0\to \pi^0\rho^0$ is not well reproduced.
The four different solutions of SU(3) breaking in $W$-exchange amplitudes here can be discriminated using the SCS mode $D^0\to \eta\phi$.  It turns out that solution (i) is preferred for both (F4) and (F1').

Topological amplitude decompositions for SCS $D\to V\!P$ decays are given in Table IV of Ref.~\cite{Cheng:2021}. As stressed in Refs.~\cite{Cheng:2019} and~\cite{Cheng:2021}, the consideration of SU(3) breaking in $T_{V,P}$ and $C_{V,P}$ alone would render even larger deviations from the data. That is why we focus only on SU(3) breaking in the $W$-exchange amplitudes for SCS $D^0$ decays. As for SCS $V\!P$ decays of $D^+$ and $D_s^+$, we have found a rule of thumb: It is necessary to consider the SU(3)-breaking effects if only one of the $T_{V,P}$ and $C_{V,P}$ topological amplitudes appears in the decay amplitude~\cite{Cheng:2021}.

Based on solutions (F4) and (F1'), the calculated branching fractions of SCS $D\to V\!P$ decays are displayed in Table~\ref{tab:SCSVP}. We note in passing that none of the solutions (F1')-(F5') can accommodate all the data of SCS channels $D^{0,+}\to\pi^{0.+}\omega$, $D^{0,+}\to \pi^{0,+}\rho^0$ and $D^0\to\eta\omega$.
Nevertheless, (F1') is the best solution among the (F') set as it accommodates the data of $D^0\to \pi^0\rho^0, \pi^0\omega$ and $\eta\omega$, although its prediction of $\B(D^+\to \pi^+\omega)$ is too large compared to the experiment, as shown in Table~\ref{tab:SCSVP}. For this reason, we compare the predictions of $D\to V\!P$ based on solutions (F4) and (F1').

\subsubsection{Doubly-Cabibbo-suppressed $D\to V\!P$ decays}
Topological amplitude decompositions for DCS $D\to VP$ decays are listed in Table~\ref{tab:DCSVP}, where we have set $T_{V,P}''=T_{V,P}$ and $C_{V,P}''=C_{V,P}$. In the table, we have also introduced $\tilde E''_V$ and $\tilde E''_P$ for reasons to be discussed below.
When setting the double-primed $W$-exchange amplitudes to be the same as the unprimed amplitudes, we find some inconsistency with the experiment, as we are going to describe below.

\begin{table}[t]
\caption{Topological amplitude decompositions, experimental and predicted branching fractions for DCS $D \to VP$ decays.  For $D ^+\to VP$ decays we have set $A_{V,P}''=A_{V,P}$. All branching fractions are quoted in units of $10^{-4}$.
\label{tab:DCSVP}
}
\medskip
\begin{ruledtabular}
\begin{tabular}{l  c c c c  }
Mode & Amplitude & ${\cal B}_{\rm expt}$~\cite{PDG}
& ${\cal B}_{\rm theo}$ (F4)  & ${\cal B}_{\rm theo}$ (F1')  \\
\hline
$D^0\to K^{*+}\,\pi^-$          & $\lambda_{ds}(T_P + E''_V)$ & $3.39^{+1.80}_{-1.02}$
&$3.54\pm0.28$   & $3.46\pm0.17$  \\
$D^0\to K^+\,\rho^-$                      & $\lambda_{ds}(T_V + E''_P)$ &
&$1.30 \pm 0.07$         &$1.32 \pm 0.04$        		
\\
$D^0\to {K}^{*0}\,\pi^0$             &$\frac{1}{\sqrt{2}}\lambda_{ds}(C_P - E''_V)$ &
&$0.84\pm0.04$       &$0.48\pm0.02$  \\           	
$D^0\to {K}^0\,\rho^0$              & $\frac{1}{\sqrt{2}}\lambda_{ds}(C_V - E''_P)$ &
& $0.25\pm0.02$        & $0.27\pm0.01$
\\
$D^0\to {K}^{*0}\,\eta$               & $\lambda_{ds}\left[ {1\over \sqrt{2}}(C_P + E''_V)\cos\phi - E''_P\sin\phi\, \right]$ &
&$0.34\pm0.02$    &$0.20\pm0.01$
\\
$D^0\to {K}^{*0}\,\eta\,'$            &$\lambda_{ds}\left[ {1\over \sqrt{2}}(C_P + E''_V)\sin\phi + E''_P\cos\phi\, \right]$ &
&$0.0019\pm0.0001$    &$0.0016\pm0.0001$  \\
$D^0\to {K}^0\,\omega$             & $-\frac{1}{\sqrt{2}}\lambda_{ds}(C_V + \tilde E''_P)$ &
&$0.66 \pm 0.03$   &$0.51 \pm 0.02$   \\
$D^0\to {K}^0\,\phi$                   &$-\lambda_{ds} \tilde E''_V$ &
&$0.22\pm0.01$    &$0.05\pm0.01$  \\
\hline
$D^+\to K^{*0}\,\pi^+$ & $\lambda_{ds}(C_P+A''_V)$ & $3.45 \pm 0.60$ & $2.52\pm0.07$ & $2.51\pm0.06$ \\
$D^+\to K^{*+}\,\pi^0$ & $\frac{1}{\sqrt{2}}\lambda_{ds}\left( T_P-A''_V \right)$ & $3.4\pm1.4$ &  $4.17\pm0.15$ & $4.12\pm0.14$ \\
$D^+\to K^0\rho^+$ & $\lambda_{ds}(C_V+A''_P)$ & & $1.40\pm0.06$  & $1.38\pm0.05$\\
$D^+\to K^+\rho^0$ & $\frac{1}{\sqrt{2}}\lambda_{ds}(T_V-A''_P)$
& $1.9 \pm 0.5$ & $1.84\pm0.05$ & $1.54\pm0.04$ \\
$D^+\to K^{*+}\,\eta$ & $\lambda_{ds}\left[{1\over\sqrt{2}}(T_P+A''_V)\cos\phi-A''_P\sin\phi\right]$  & $$ & $1.44\pm0.05$ & $1.29\pm0.04$ \\
$D^+\to K^{*+}\,\eta^{\prime}$ & $\lambda_{ds}\left[{1\over\sqrt{2}}(T_P+A''_V)\sin\phi+A''_P\cos\phi\right]$
&  & $0.016\pm0.001$ & $0.021\pm0.001$ \\
$D^+\to K^+\omega$ & $\frac{1}{\sqrt{2}}\lambda_{ds}(T_V+A''_P)$ &  $0.57^{+0.25}_{-0.21}$ & $2.09\pm0.05$ & $2.42\pm0.05$ \\
$D^+\to K^+\phi$ & $\lambda_{ds}A''_V$ & $0.090\pm0.012$ & $0.057\pm0.002$  & $0.032\pm0.003$ \\
\hline
$D_s^+\to K^{*+}\,K^0$ & $\lambda_{ds}(T_P+C_V)$ &
& $1.47 \pm 0.09$ & $1.44 \pm 0.11$  \\
$D_s^+\to K^{*0}\,K^+$ & $\lambda_{ds}(T_V+C_P)$ & $0.90\pm0.51$
& $0.20 \pm 0.02$ & $0.11 \pm 0.02$  \\
\end{tabular}
\end{ruledtabular}
\end{table}

\begin{table}[t]
\caption{Strong phase difference $\delta_n$ (in units of degrees) between $D^0\to n$ and $D^0\to \bar n$ evaluated using solutions (F4) and (F1') for topological amplitudes in $D\to V\!P$ decays.
\label{tab:phase}
}
\vskip 0.2 cm
\begin{ruledtabular}
\begin{tabular}{l rrrr l rrrr }
Solution & \multicolumn{2}{c}{F4} & \multicolumn{2}{c}{F1'} & Solution
& \multicolumn{2}{c}{F4} & \multicolumn{2}{c}{F1'}
\\
\cline{2-3} \cline{4-5} \cline{7-8} \cline{9-10}
Mode & $|\delta_n|$ &  $\cos\delta_n$ & $|\delta_n|$ & $\cos\delta_n$
& ~~Mode & $|\delta_n|$ &  $\cos\delta_n$ & $|\delta_n|$ & $\cos\delta_n$
\\
\hline
$K^{*-}\pi^+$  & 3.1 & 0.999 & 2.1 & 0.999 &
~~$K^-\rho^+$    & 22 & 0.927 & 28 & 0.886   \\
$\ov {K}^{*0}\pi^0$  & 3.0 & 0.999 & 16 & 0.959 &
~~$\ov K^0\rho^0$    & 2.7 & $0.999$ & 26 & 0.897    \\
$\ov {K}^{*0}\eta$  & 6.9 & $0.993$ & 38 & 0.787 &
~~$\ov K^{*0}\eta'$    & 5.4 & 0.996 & 1.5 & 1   \\
$\ov {K}^{0}\omega$  & 6.7 & 0.993 & 18 & 0.949 &
~~$\ov K^{0}\phi$    & 10 & $0.984$ & 31 & 0.856   \\
$\rho^-\pi^+$  & $\sim 0$ & 1 & 2.3 & 0.999 &
~~$\rho^+\pi^-$  & $\sim 0$ & 1 & 2.3 & 0.999   \\
$\ov K^{*0}K^{0}$    & $4.3$ & 0.997 & $7.3$ & $0.992$   &
~~$K^{*-}K^+$    & $7.2$ & 0.992 & 33 & 0.839   \\
\end{tabular}
\end{ruledtabular}
\end{table}

In order to evaluate the $D^0$-$\ov D^0$ mixing parameter $y$ to be discussed in Sec.~\ref{sec:y}, we need to know the strong phase difference $\delta_n$ between the $D^0\to n$ and $D^0\to\bar n$ [see Eq.~(\ref{eq:y}) below]. In principle, this can be evaluated in the TDA. For example,
the following $D\to V\!P$ modes have the expressions
\be \label{eq:CFDCSdecays}
\begin{split}
A(D^0\to K^{*-}\pi^+)=\lambda_{sd}(T_V+E_P), &\qquad A(D^0\to K^{*+}\pi^-)=\lambda_{ds}(T_P+E''_V),
\\
A(D^0\to K^{-}\rho^+)=\lambda_{sd}(T_P+E_V), &\qquad A(D^0\to K^{+}\rho^-)=\lambda_{ds}(T_V+E''_P),
\\
A(D^0\to \ov K^{*0}\pi^0)={1\over\sqrt{2}}\lambda_{sd}(C_P-E_P), &\qquad A(D^0\to K^{*0}\pi^0)={1\over\sqrt{2}}\lambda_{ds}(C_P-E''_V),
\\
A(D^0\to \ov K^{0}\phi)=\lambda_{sd}E_P, &\qquad A(D^0\to K^{0}\phi)=\lambda_{ds}\tilde E''_V,
\\
A(D^0\to \rho^-\pi^+)=\lambda_d(T_V+E_P^d), &\qquad A(D^0\to \rho^+\pi^-)=\lambda_d(T_P+E_V^d),
\\
A(D^0\to K^{*-}K^+)=\lambda_s(T_V+E_P^s), &\qquad A(D^0\to K^{*+}K^-)=\lambda_s(T_P+E_V^s),
\end{split}
\en
in the diagrammatic approach, where $E_{V,P}^{d,s}$ are introduced before in Eq. (\ref{eq:SU(3)inW}) to account for SU(3) violation in the $W$-exchange amplitudes $E_V$ and $E_P$ appearing in SCS decays.
The strong phases are then given by
\be
\begin{split}
\delta_{K^{*+}\pi^-}={\rm arg}[(T_P+E''_V)/(T_V+E_P)], &\qquad
\delta_{K^{+}\rho^-}={\rm arg}[(T_V+E''_P)/(T_P+E_V)],
\\
\delta_{K^{*0}\pi^0}={\rm arg}[(C_P-E''_V)/(C_P-E_P)], &\qquad
\delta_{K^{0}\phi}={\rm arg}[\tilde E''_V/E_P],
\\
\delta_{\rho^+\pi^-}={\rm arg}[(T_P+E_V^d)/(T_V+E_P^d)], &\qquad
\delta_{K^{*+}K^-}={\rm arg}[(T_P+E_V^s)/(T_V+E_P^s)].
\end{split}
\en

In the TDA, a fit to the CF $D\to VP$ modes indicates that $E_V$ and $E_P$ differ in both magnitude and phase. For example, in solution (F4) we have (see Table~\ref{tab:CFVPsolution_1})
\be
E_V=(1.14\pm0.04)e^{i(257\pm2)^\circ}, \qquad E_P=(1.66\pm0.03)e^{i(107\pm3)^\circ},
\en
in units of $10^{-6}(\epsilon\cdot p_D)$. When setting $E''_V=E_V$ and $E''_P=E_P$, we find that $\delta_{K^{*+}\pi^-}=\delta_{K^{+}\rho^-}=48^\circ$ and $\delta_{K^{*0}\pi^0}=72^\circ$. They are too large compared to the experimental values of $|\delta_{K^{*+}\pi^-}|= (6.1\pm0.7)^\circ$ obtained by Belle~\cite{Belle:D0Kspipi} and $(2.4\pm1.1)^\circ$ by BaBar~\cite{BaBar:D0Kspipi} both extracted from $D^0\to K_S^0\pi^+\pi^-$ decays~\footnote{The relative phase difference was measured to be $(173.9\pm0.7)^\circ$ by Belle~\cite{Belle:D0Kspipi} and $(177.6\pm1.1)^\circ$ by BaBar~\cite{BaBar:D0Kspipi}.  These results are close to the $180^\circ$ expected from Cabibbo factors, i.e., the relative minus sign between $\lambda_{sd}$ and $\lambda_{ds}$.} and $\delta_{K\pi\pi^0}=(18\pm10)^\circ$ extracted from $D^0\to K^-\pi^+\pi^0$~\cite{PDG}. To overcome this difficulty, we shall consider the SU(3) breaking effects in $W$-exchange amplitudes in the DCS $D\to V\!P$ sector such that $T_P+E''_V$ in $D^0\to K^{*-}\pi^+$ is almost parallel to $T_V+E_P$ in $D^0\to K^{*+}\pi^-$. This can be achieved by having
\be \label{eq:EVpp}
E''_V=
\begin{cases}
1.4e^{i 220^\circ} E_V, & {\rm solution~(F4),}
\\
0.8 e^{-i 230^\circ} E_V, & {\rm solution~(F1')}.
\end{cases}
\en
and
\be  \label{eq:EPpp}
E''_P=
\begin{cases}
0.6  e^{i 140^\circ} E_P, & {\rm solution~(F4),}
\\
0.2 e^{i 230^\circ} E_P, & {\rm solution~(F1')}.
\end{cases}
\en

The results of $\delta_n$'s evaluated using solutions (F4) and (F1') for topological amplitudes are summarized in Table~\ref{tab:phase}. CLEO has analyzed the decays $D^0\to K_S^0K^-\pi^+$, $D^0\to K_S^0K^+\pi^-$ and obtained the phase $\delta_{K^*K}=(-16.6\pm18.4)^\circ$~\cite{CLEO:KSKpi}. Our results for $\delta_{K^{*0}K^0}$ and $\delta_{K^{*-}K^+}$ are consistent with the experiment. Our predictions of $\cos\delta_n$ for those $D\to V\!P$ modes which are not \CP eigenstates are generally different from that calculated in the FAT approach~\cite{Jiang} where $\cos\delta_n$ is close to unity.

\begin{table}[tp!]
\caption{Topological amplitude decompositions and branching fractions (in units of \%) for $D\to K_{S,L}^0V$ decays. Experimental results are taken from Ref.~\cite{PDG}. Predictions based on the FAT approach~\cite{Wang} are included for comparison.
  \label{tab:VPKLKS}  }
\medskip
\scriptsize{
\begin{ruledtabular}
\begin{tabular}{l  c c c c  c  }
  Mode~~~ & Representation & ${\cal B}_{\rm expy}$ &  ${\cal B}_{\rm theo}$ (F4) &   ${\cal B}_{\rm theo}$ (F1') &  ${\cal B}_{\rm theo}$ (FAT) \\
\hline
  $D^0\to K_S^0\rho^0$ &  ${1\over 2}[\lambda_{sd}(C_V-E_V)-\lambda_{ds}(C_V-E_P'')]$ &  $0.63^{+0.06}_{-0.08}$  &  $0.68\pm0.05$  &  $0.68\pm0.05$ & $0.50\pm0.11$  \\
  $D^0\to K_L^0\rho^0$ & ${1\over 2}[\lambda_{sd}(C_V-E_V)+\lambda_{ds}(C_V-E_P'')]$ &  $$  &  $0.57\pm0.04$ & $0.58\pm0.04$ & $0.40\pm0.09$ \\
  $D^0\to K_S^0\omega$ & ${1\over 2}[\lambda_{sd}(C_V+E_V)-\lambda_{ds}(C_V+\tilde E_P'')]$ & $1.11\pm0.06$ &   $1.27\pm0.06$   & $1.25\pm0.06$ & $1.18\pm0.19$  \\
  $D^0\to K_L^0\omega$ & ${1\over 2}[\lambda_{sd}(C_V+E_V)+\lambda_{ds}(C_V+\tilde E_P'')]$ & $1.16\pm0.04$ & $1.03\pm0.05$  & $1.05\pm0.05$ & $0.95\pm0.15$ \\
  $D^0\to K_S^0\phi$ & ${1\over\sqrt{2}}(\lambda_{sd}E_P-\lambda_{ds}\tilde E_V'')$ & $0.413\pm0.031$ &   $0.458\pm0.018$    & $0.431\pm0.017$  & $0.40\pm0.04$ \\
  $D^0\to K_L^0\phi$ & ${1\over\sqrt{2}}(\lambda_{sd}E_P+\lambda_{ds}\tilde E_V'')$ & $0.414\pm0.023$ &  $0.374\pm0.016$ & $0.397\pm0.017$  & $0.33\pm0.03$ \\
  $D^+\to K_S^0\rho^+$ &  ${1\over \sqrt{2}}[\lambda_{sd}(T_P+C_V)-\lambda_{ds}(C_V+A_P)]$ &  $6.14^{+0.60}_{-0.35}$   & $6.38\pm0.44$  & $6.26\pm0.52$ & $4.99\pm0.50$ \\
   $D^+\to K_L^0\rho^+$ &   ${1\over \sqrt{2}}[\lambda_{sd}(T_P+C_V)+\lambda_{ds}(C_V+A_P)]$ &  $$  & $7.19\pm0.44$  & $7.03\pm0.52$ & $5.37\pm0.50$ \\
  $D_s^+\to K_S^0K^{*+}$ &  ${1\over \sqrt{2}}[\lambda_{sd}(C_V+A_P)-\lambda_{ds}(T_P+C_V)]$ &  $0.77\pm0.07$  &  $0.79\pm0.04$ & $0.78\pm0.04$ &$1.20\pm0.36$  \\
   $D_s^+\to K_L^0K^{*+}$ &  ${1\over \sqrt{2}}[\lambda_{sd}(C_V+A_P)+\lambda_{ds}(T_P+C_V)]$ &  $$ & $1.09\pm0.04$  & $1.07\pm0.04$ & $1.37\pm0.33$ \\
\end{tabular}
\end{ruledtabular}}
\end{table}
%

Considering the possible resonant contributions of $D^0\to K^{*+}\pi^-$, $K^{*0}\pi^0$ and $K^+\rho^-$ to the DCS decay $D^0\to K^+\pi^-\pi^0$, our prediction of $\B(D^0\to K^+\rho^-)\sim 1.30\times 10^{-4}$ (see Table~\ref{tab:DCSVP}) is consistent with $\B(D^0\to K^+\pi^-\pi^0)=(3.13^{+0.60}_{-0.56}\pm0.15)\times 10^{-4}$ measured by BESIII~\cite{BESIII:D0Kppi-pi0}.

Using the double-primed amplitudes given in Eqs.~(\ref{eq:EVpp}) and (\ref{eq:EPpp}) and assuming that $\tilde E''_{V,P}=E''_{V,P}$, the calculated $D\to K_{S,L}^0V$ decay rates and their asymmetries $R(D,V)$ defined in analogue to Eq.~(\ref{eq:R}) are shown in Tables~\ref{tab:VPKLKS} and \ref{tab:RVP}, respectively. We predict that $\B(D^0\to K_S^0V)>\B(D^0\to K_L^0V)$ for $V=\rho^0,\omega,\phi$, $\B(D^+\to K_S^0\rho^+)<\B(D^+\to K_L^0\rho^+)$ and $\B(D_s^+\to K_S^0 K^{*+})<\B(D_s^+\to K_L^0 K^{*+})$. Experimentally, we see that $\B(D^0\to K_S^0\omega)\gsim\B(D^0\to K_L^0\omega)$ and  $\B(D^0\to K_S^0\phi)\approx\B(D^0\to K_L^0\phi)$. The last one implies that $\tilde E_V''$ is nearly orthogonal to $E_P$ rather than that given by Eq.~(\ref{eq:EVpp}). Indeed, this can be achieved by letting $\tilde E_V''=E_V e^{i300^\circ}$ using solution (F4), which leads to $\B(D^0\to K_{S,L}^0\phi)\approx 0.415\%$. However, if we set $E_V''=E_V e^{i300^\circ}$, we will have $\delta_{K^{*+}\pi^-}=32^\circ$ and $\B(D^0\to K^{*+}\pi^-)=1.06\times 10^{-4}$, both not consistent with the experiment.  In the spirit of the TDA, there should be only one double-primed amplitude for $W$-exchange; that is, one should have $\tilde E_V''=E_V''$. It is conceivable that $D^0\to K^0\omega$ and $K^0\phi$ decays receive additional singlet contributions $S^\omega$ and $S^\phi$~\cite{Cheng:2011qh}, respectively, owing to the SU(3)-singlet nature of the vector mesons $\omega$ and $\phi$. We shall leave this and the above-mentioned issues to a future study.

\begin{table}[t]
\caption{$K_S^0-K_L^0$ asymmetries for $D\to K_{S,L}^0V$ decays. Experimental measurements are taken from Ref.~\cite{BESIII:D0KL}.
  \label{tab:RVP}  }
  \medskip
\begin{ruledtabular}
\begin{tabular}{l  c c c c  c  }
   & $R(D^0,\rho^0)$ & $R(D^0,\omega)$ &  $R(D^0,\phi)$ &   $R(D^+,\rho^+)$ &  $R(D_s^+,K^{*+})$   \\
\hline
   (F4) &  $0.090\pm0.052$ & $0.106\pm0.034$  &  $0.101\pm0.029$ & $-0.060\pm0.046$ & $-0.164\pm0.032$  \\
    (F1') &  $0.083\pm0.050$ & $0.089\pm0.035$  &  $0.041\pm0.029$ & $-0.058\pm0.055$ & $-0.159\pm0.028$  \\
    Expt &  -- & $-0.024\pm0.031$  &  $-0.001\pm0.047$ & -- & --  \\
\end{tabular}
\end{ruledtabular}
\end{table}
%

In the FAT approach~\cite{Wang}, the double-primed topological amplitudes are taken to be the same as the unprimed ones; that is, $E_V''=E_V$ and $E_P''=E_P$. Moreover, it is assumed that $E_P=E_V$. Consequently, this assumption leads to~\cite{Wang}
\be
R(D^0,\rho^0)=R(D^0,\omega)=R(D^0,\phi)=2\tan^2\theta_C=0.107\,.
\en
Obviously, the predicted $K_S^0-K_L^0$ asymmetries $R(D^0,\omega)$ and $R(D^0,\phi)$ are wrong in sign (see Table~\ref{tab:RVP}).

\section{$D\to VV$ decays \label{sec:DtoVV}}

\begin{table}[t]
\caption{Topological amplitude decompositions of CF, SCS and DCS $D\to VV$ decays. The subscript $h$ denotes the helicity state, or the spin state in the trasnversity basis, or the partial-wave amplitude. Here $\lambda_{sd}\equiv V_{cs}^*V_{ud}$, $\lambda_{ds}\equiv V_{cd}^*V_{us}$, $\lambda_{d}\equiv V_{cd}^*V_{ud}$ and $\lambda_{s}\equiv V_{cs}^*V_{us}$.
  \label{tab:AmpDtoVV}  }
\medskip
\begin{center}
\begin{tabular}{l  c  | l c  }
\hline\hline
  Mode~~~ & ~~~~Representation~~~~~~ & ~~~Mode & Representation \\
\hline
  $D^0\to K^{*-}\rho^+$ & $\lambda_{sd}(T_h+E_h)$ & ~~~$D^0\to \ov K^{*0}\rho^0$~~~ & ${1\over\sqrt{2}}\lambda_{sd}(C_h-E_h)$ \\
  $D^0\to \ov K^{*0}\omega$ & ${1\over\sqrt{2}}\lambda_{sd}(C_h+E_h)$ & ~~~$D^0\to \ov K^{*0}\phi$
  & $\lambda_{sd}E_h$\\
  $D^0\to \rho^{+}\rho^-$ & $\lambda_d(T_h+E^{d}_h)$ & ~~~$D^0\to \rho^{0}\rho^0$~~~ & ${1\over \sqrt{2}}\lambda_d(C_h-E^{ d}_h)$ \\
  $D^0\to K^{*+}K^{*-}$ & $\lambda_s(T_h+E^{s}_h)$ & ~~~$D^0\to K^{*0}\ov K^{*0}$~~~ & $\lambda_{d}E^{d}_h+\lambda_{s}E^{s}_h$ \\
  $D^0\to \rho^{0}\omega$ & $-\lambda_d E^{d}_h$ & ~~~$D^0\to \rho^0\phi$~~~ & ${1\over\sqrt{2}}\lambda_s C_h$ \\
  $D^0\to \omega\omega$ & ${1\over\sqrt{2}}\lambda_d(C_h+E^{d}_h)$ & ~~~$D^0\to \omega\phi$~~~ & ${1\over\sqrt{2}}\lambda_s C_h$ \\
  $D^0\to \phi\phi$~~~ & $\sqrt{2}\lambda_s E_h$ \\
  $D^0\to K^{*+}\rho^-$ & $\lambda_{ds}(T_h+E_h)$ & ~~~$D^0\to K^{*0}\rho^0$~~~ & ${1\over\sqrt{2}}\lambda_{ds}(C_h-E_h)$ \\
  $D^0\to K^{*0}\omega$ & ${1\over\sqrt{2}}\lambda_{ds}(C_h+E_h)$ & ~~~$D^0\to K^{*0}\phi$
  & $\lambda_{ds}E_h$\\
  \hline
  $D^+\to \ov K^{*0}\rho^+$ & $\lambda_{sd}(T_h+C_h)$ & &  \\
  $D^+\to \rho^+\rho^0$ & ${1\over\sqrt{2}}\lambda_d(T_h+C_h)$ & ~~~$D^+\to K^{*+}\ov K^{*0}$~~~ & $\lambda_{s}T_h+\lambda_d A_h$\\
  $D^+\to \rho^+\omega$ & ${1\over \sqrt{2}}\lambda_d(T_h+C_h+2A_h)$  & ~~~$D^+\to \rho^+\phi$~~~ & $\lambda_s C_h$  \\
  $D^+\to K^{*0}\rho^+$ & $\lambda_{ds}(C_h+A_h)$ &   ~~~$D^+\to K^{*+}\phi$ & $\lambda_{ds}A_h$ \\
  $D^+\to K^{*+}\rho^0$ & ${1\over\sqrt{2}}\lambda_{ds}(T_h-A_h)$ &  $~~~D^+\to K^{*+}\omega$ & ${1\over\sqrt{2}}\lambda_{ds}(T_h+A_h)$ \\
  \hline
  $D_s^+\to \ov K^{*0}K^{*+}$ & $\lambda_{sd}(C_h+A_h)$ & ~~~$D_s^+\to \rho^+\rho^0$ & 0 \\
  $D_s^+\to\rho^+\omega$ & ${1\over\sqrt{2}}\lambda_{sd}\,A_h$ & ~~~$D_s^+\to\rho^+\phi$ & $\lambda_{sd} T_h$  \\
  $D_s^+\to\rho^+ K^{*0}$ & $\lambda_d T_h+\lambda_s A_h$ & ~~~$D_s^+\to\rho^0 K^{*+}$ & ${1\over\sqrt{2}}(\lambda_d\, C_h-\lambda_{s} A_h)$  \\
  $D_s^+\to K^{*+}\omega$ & ~~${1\over\sqrt{2}}(\lambda_d C_h+\lambda_s A_h)$~~ & ~~~$D_s^+\to K^{*+}\phi$ & $\lambda_s(T_h+C_h+A_h)$  \\
  $D_s^+\to K^{*+}K^{*0}$ & $\lambda_{ds}(T_h+C_h)$ \\
   \hline\hline
\end{tabular}
\end{center}
\end{table}
%

The underlying mechanism for $D\to VV$ decays is more complicated than $PV$ and $PP$ modes as each $V$ involves three polarization vectors. In general, the decay amplitudes can be expressed in several different but equivalent bases.  The helicity amplitudes $H_0, H_+$ and $H_-$ can be related to the spin amplitudes in the transversity basis $(A_0, A_\|, A_\bot)$ defined in terms of the linear polarization of the vector mesons, or to the partial-wave amplitudes $(S,P,D)$ via:
\be \label{eq:Ampbases}
\begin{split}
A_0 &= H_{0}= -{1\over\sqrt{3}}\, S+\sqrt{2\over 3}\, D ~,
\\
A_\| &= {1\over\sqrt{2}}(H_{+}+ H_{-})=\sqrt{2\over 3}
\,S+{1\over\sqrt{3}}\,D ~,
\\
A_\bot &= {1\over\sqrt{2}}(H_{+}- H_{-})=P ~,
\end{split}
\en
or
\be
\begin{split}
S &= {1\over\sqrt{3}}(-A_0+\sqrt{2}A_\|)={1\over\sqrt{3}}(-H_0+H_++H_-),
\\
P &= A_\bot={1\over\sqrt{2}}(H_+-H_-),
\\
D &= {1\over\sqrt{3}}(\sqrt{2}A_0+A_\|)={1\over\sqrt{6}}(2H_0+H_++H_-),
\end{split}
\en
where we have followed the sign convention of Ref.~\cite{Dighe}. The decomposition of
topological diagram amplitudes of CF, SCS and DCS $D\to VV$ decays is collected in Table~\ref{tab:AmpDtoVV}. Note that although the decays $D^0\to \ov K^{*0}\phi$, $K^{*0}\phi$ and $D^+\to K^{*+}\phi$ are kinematically prohibited, they can proceed through the finite width of $K^*(892)$. Indeed,  $D^0\to \ov K^{*0}\phi$ has been observed in the four-body decay $D^0\to K^-K^-K^+\pi^+$~\cite{FOCUS:2003gcs}. The decay $D^0\to \phi\phi$ is also kinematically disallowed, but we include it in Table~\ref{tab:AmpDtoVV} in order to show that the mixing parameter $y_{_{VV}}$ to be discussed in Sec.~\ref{sec:y} vanishes in the SU(3) limit.

For charmless $B\to VV$ decays, it is na{\"i}vely expected that the helicity amplitudes $H_h$  respect the hierarchical pattern $H_0:H_-: H_+=1:(\Lambda_{\rm QCD}/ m_b):(\Lambda_{\rm QCD}/m_b)^2$. Hence, they are expected to be dominated by the longitudinal polarization states and satisfy the scaling law,
\begin{eqnarray} \label{eq:scaling}
1-f_L={\cal O}\left({m^2_V\over m^2_B}\right),
\end{eqnarray}
with
\be
f_L\equiv \frac{\Gamma_L}{\Gamma}
=\frac{| A_0|^2}{|A_0|^2+|A_\parallel|^2+|A_\bot|^2}=\frac{| H_0|^2}{|H_0|^2+|H_+|^2+|H_-|^2}.
\label{eq:fL}
\en
This prediction has been confirmed in the tree-dominated $B$ decays such as $B^0\to \rho^+\rho^-$
and $B^+\to \rho^+\rho^0$. However, the large fraction of transverse polarization observed in the penguin-doiminated decays
$B\to\phi K^*$, $B^+\to \omega K^{*+}$ and $B\to K^*\rho$ (except $B^+\to K^{*+}\rho^0$)  decays~\cite{PDG} is a surprise and poses an interesting challenge for
theoretical interpretations. In $D\to VV$ decays, we shall see that na{\"i}ve factorization leads to the prediction that $f_L$ is comparable to or smaller than the transverse polarization.

Under factorization, the factorizable matrix element for the $D\to V_1V_2$ decay reads
\be
\begin{split}
X_h^{(DV_1,V_2)} \equiv & \la V_2 |J^{\mu}|0\ra\la
V_1|J'_{\mu}|D \ra
\\
=&
- if_{V_2}m_{V_2}\Bigg[
(\vp^*_1\cdot\vp^*_2) (m_{D}+m_{V_1})A_1^{ DV_1}(m_{V_2}^2)
\\
& \qquad
- (\vp^*_1\cdot p_{_{D}})(\vp^*_2 \cdot p_{_{D}})
\frac{2A_2^{DV_1}(m_{V_2}^2)}{m_{D}+m_{V_1}} +
i\epsilon_{\mu\nu\alpha\beta}\vp^{*\mu}_2\vp^{*\nu}_1p^\alpha_{_{D}}
p^\beta_1 \frac{2V^{ DV_1}(m_{V_2}^2)}{m_{D}+m_{V_1}} \Bigg] ,
\end{split}
\en
where use of the conventional definition for form factors~\cite{BSW} has been made.  The longitudinal ($h=0$) and transverse ($h=\pm$) components of $X^{(DV_1,V_2)}_h$ are given by
 \be \label{eq:Xh}
 \begin{split}
 X_0^{(DV_1,V_2)} &= {if_{V_2}\over 2m_{V_1}}\left[
 (m_D^2-m_{V_1}^2-m_{V_2}^2)(m_D+m_{V_1})A_1^{DV_1}(q^2)-{4m_D^2p_c^2\over
 m_D+m_{V_1}}A_2^{DV_1}(q^2)\right] ,
 \\
 X_\pm^{(DV_1,V_2)} &= -if_{V_2}m_Dm_{V_2}\left[
 \left(1+{m_{V_1}\over m_D}\right)A_1^{DV_1}(q^2)\mp {2p_c\over
 m_D+m_{V_1}}V^{DV_1}(q^2)\right] .
 \end{split}
 \en
We see from Eq.~(\ref{eq:Ampbases}) that the amplitude $A_\parallel$ is governed by the form factor $A_1^{DV_1}$, while $A_\bot$ is related to $V^{DV_1}$. The decay rate reads
\be
\begin{split}
\Gamma(D\to V_1V_2)
&= {p_c\over 8\pi m_D^2} \left( |H_{0}|^2+|H_{+}|^2+|H_{-}|^2 \right) ,
\\
&= {p_c\over 8\pi m_D^2} \left( |A_{0}|^2+|A_\bot|^2+|A_\||^2 \right) ,
\\
&= {p_c\over 8\pi m_D^2} \left( |S|^2+|P|^2+|D|^2 \right) .
\end{split}
\en
%

\begin{table}[tp!]
\caption{Experimental results for the branching fractions of $D^0\to VV$ in partial waves.  Data are taken from Ref.~\cite{PDG} unless specified otherwise.
\label{tab:DVVdata}}
\vspace{6pt}
\footnotesize{
\begin{ruledtabular}
\begin{tabular}{l l c c c c  c }
Meson & Mode & $S$-wave &  $P$-wave & $D$-wave & $\B_{\rm expt}$  \\
\hline
$D^0$ &
$K^{*-} \rho^+$ & $(1.4\pm 0.4)\%$ \footnotemark[1]  & $(0.9\pm 0.2)\%$ & $(2.9\pm 0.8)\%$  & $$ \\
& $\ov K^{*0} \rho^0$ & $(8.0\pm 1.2)\times 10^{-3}$ \footnotemark[2] & $(2.8\pm 0.3)\times 10^{-3}$ & $(9.8\pm 1.0)\times 10^{-3}$  & $(1.52\pm0.08)\%$ \\
& & $(6.0\pm 0.4)\times 10^{-3}$ \footnotemark[3] & $(5.0\pm 0.2)\times 10^{-3}$ & $(7.0\pm 0.6)\times 10^{-3}$  &  \\
& $\ov K^{*0} \omega$ &  & $$ & $$ & $(1.1\pm0.5)\%$ \\
& $\ov K^{*0}\phi$ & & & & $(3.30\pm0.64)\times 10^{-4}$ \\
& $\rho^+\rho^-$ & $(1.2\pm0.4)\times 10^{-3}$ \footnotemark[4] & $(1.8\pm0.3)\times 10^{-3}$ & $(3.3\pm0.5)\times 10^{-3}$ & $(7.81\pm1.14)\times 10^{-3}$ \\
& $\rho^0\rho^0$ & $(0.8\pm0.4)\times 10^{-4}$ \footnotemark[5] & $(4.6\pm0.7)\times 10^{-4}$ & $(1.1\pm0.2)\times 10^{-3}$ & $(1.33\pm0.21)\times 10^{-3}$ \\
& $\rho^0\rho^0$ & $(1.8\pm1.3)\times 10^{-4}$ \footnotemark[6] & $(5.3\pm1.3)\times 10^{-4}$ & $(6.2\pm3.0)\times 10^{-4}$ & $(1.33\pm0.35)\times 10^{-3}$ \\
& $\rho^0\phi$ & $(1.4\pm0.1)\times 10^{-3}$ & $(8.1\pm3.9)\times 10^{-5}$ & $(8.5\pm2.8)\times 10^{-5}$ & $(1.56\pm0.13)\times 10^{-3}$ \\
& $\omega\phi$ & $$ & $$ & $$ & $(6.48\pm1.04)\times 10^{-4}$ \\
& $K^{*0}\ov K^{*0}$ & $(5.04\pm0.29)\times 10^{-4}$ & $(2.70\pm0.18)\times 10^{-4}$ & $(1.06\pm0.09)\times 10^{-4}$ & $(0.88\pm0.04)\times 10^{-3}$\\
\hline
$D_s^+$ & $K^{*+}\ov K^{*0}$ & $(5.01\pm 0.92)\%$ \footnotemark[7] & $(1.10\pm0.19)\%$ & $(0.65\pm0.16)\%$ &  $(5.93\pm0.88)\%$ \\
& & $(3.96\pm 0.26)\%$ \footnotemark[8] & $(1.67\pm0.16)\%$ & $(0.81\pm0.14)\%$ &  $(5.64\pm0.35)\%$ \\
& $\rho^+\phi$ & $(4.27\pm 0.32)\%$ \footnotemark[8]  & $(1.06\pm0.11)\%$ & $(0.37\pm0.09)\%$ & $(5.59\pm0.34)\%$ \\
& $K^{*0} \rho^+$ & $(1.41\pm 0.24)\times 10^{-3}$ \footnotemark[9]  & $(2.53\pm0.31)\times 10^{-3}$ & $$ & $(3.95\pm0.39)\times 10^{-3}$ \\
& $K^{*+} \rho^0$ & $$   & $(0.42\pm0.17)\times 10^{-3}$ \footnotemark[9] & $$ & $$ \\
\hline
$D^+$ & $\overline{K}^{*0}\rho^+$
& $(5.52\pm0.55)\%$ \footnotemark[10] & $(2.94\pm1.02)\times 10^{-3}$ &  & $(5.82\pm0.56)\%$\\

\end{tabular}
\footnotetext[1]{Partial waves are taken from the measured fit fractions in the decay $D^0\to K^-\pi^+\pi^0\pi^0$~\cite{BESIII:D0toKpi2pi0}. However, measurement of the fit fraction of $D^0\to K^{*-}\rho^+$ was not reported by BESIII. Mark~III results of $\B(D^0\to K^{*-}\rho^+)=(6.5\pm2.6)\%$
and $(3.1\pm1.2)\%$, $(3.4\pm2.0)\%$ for the longitudinal and transverse branching fractions, respectively, were listed in the 2009 version of PDG~\cite{PDG2009}.}
\footnotetext[2]{Taken from the BESIII measurement of $D^0\to K^-\pi^+\pi^+\pi^-$~\cite{BESIII:D0toK3pi}.}
\footnotetext[3]{Taken from the LHCb measurement of $D^0\to K^-\pi^+\pi^+\pi^-$~\cite{LHCb:D0toK3pi}.}
\footnotetext[4]{Taken from the BESIII measurement of $D^0\to \pi^+\pi^-\pi^0\pi^0$~\cite{BESIII:D0to4pi}.}
\footnotetext[5]{Taken from the BESIII measurement of $D^0\to \pi^+\pi^-\pi^+\pi^-$~\cite{BESIII:D0to4pi}.}
\footnotetext[6]{Partial waves are taken from Ref.~\cite{dArgent:2017gzv}. Branching fractions of $D^0\to \rho^0\rho^0$ in the transversity basis also have been measured by FOCUS~\cite{FOCUS:D0to2pip2pim}. The results read $(1.85\pm0.13)\times 10^{-3}$ for the total branching fraction and $(1.27\pm0.10)\times 10^{-3}$, $(4.8\pm0.6)\times 10^{-4}$ and $(8.3\pm3.2)\times 10^{-5}$ for the longitudinal, perpendicular and parallel components, respectively. The longitudinal polarization $f_L=0.71\pm0.04\pm0.02$ was obtained.}
\footnotetext[7]{Taken from the BESIII measurement of $D_s^+\to K_S^0K^-\pi^+\pi^+$~\cite{BESIII:DstoKSKpipi}.}
\footnotetext[8]{Taken from the BESIII measurement of $D_s^+\to K^-K^+\pi^+\pi^0$~\cite{BESIII:DstoKKpipi}.}
\footnotetext[9]{Taken from the BESIII measurement of $D_s^+\to K^+\pi^+\pi^-\pi^0$~\cite{BESIII:DstoK3pi}.}
\footnotetext[10]{Taken from the BESIII measurement of $D^+\to K_S^0\pi^+\pi^0\pi^0$~\cite{BESIII:DtoK3pi}.}
\end{ruledtabular}}
\end{table}

\begin{table}[t]
\caption{Branching fractions of CF, SCS and DCS $D^0\to VV$ decays in partial waves calculated in the factorization approach.  Data are taken from Table~\ref{tab:DVVdata}.  Since the $W$-exchange contributions are neglected in na{\"i}ve factorization, no estimate is made for the branching fractions of $K^{*0}\ov K^{*0},\ov K^{*0}\phi,K^{*0}\phi$ and $\rho^0\omega$.
\label{tab:DVV}}
\vspace{6pt}
\footnotesize{
\begin{ruledtabular}
\begin{tabular}{l c c c c c c c }
Mode & $S$-wave &  $P$-wave & $D$-wave &  $f_L$ & $\B_{\rm theo}$ & $\B_{\rm expt}$  \\
\hline
$D^0\to K^{*-} \rho^+$ & 7.1\% & $3.9\times 10^{-3}$ & $1.2\times 10^{-3}$ & 0.43 & $7.6\%$& $(6.5\pm2.5)\%$ \\
$D^0\to \ov K^{*0} \rho^0$ & 1.74\% & $1.5\times 10^{-3}$ & $2.7\times 10^{-4}$ & 0.42 & 1.9\% & $(1.52\pm0.08)\%$ \\
$D^0\to \ov K^{*0} \omega$ & 1.6\% & $1.3\times 10^{-3}$ & $2.0\times 10^{-4}$ & 0.41 & 1.8\% & $(1.1\pm0.5)\%$ \\
\hline
$D^0\to K^{*+} K^{*-}$ & $3.3\times 10^{-3}$ & $9.4\times 10^{-5}$ & $8.0\times 10^{-6}$ & 0.37 & $3.4\times 10^{-3}$ & $$ \\
$D^0\to \rho^+\rho^-$ & $3.4\times 10^{-3}$ & $3.6\times 10^{-4}$ & $1.5\times 10^{-4}$ & 0.49 &  $3.9\times 10^{-3}$ & $(7.81\pm1.14)\times 10^{-3}$ \\
$D^0\to \rho^0\rho^0$ & $0.87\times 10^{-3}$ & $0.93\times 10^{-4}$ & $3.7\times 10^{-5}$ & 0.49 & $1.00\times 10^{-3}$ & $(1.33\pm0.21)\times 10^{-3}$ \\
$D^0\to \omega\omega$ & $5.9\times 10^{-4}$ & $6.1\times 10^{-5}$ & $2.0\times 10^{-5}$ & 0.47 & $6.7\times 10^{-4}$ \\
$D^0\to \rho^0\phi$ & $6.2\times 10^{-4}$ & $2.5\times 10^{-5}$ & $1.1\times 10^{-6}$ & 0.36 & $6.5\times 10^{-4}$ & $(1.56\pm0.13)\times 10^{-3}$ \\
$D^0\to \omega\phi$ & $5.9\times 10^{-4}$ & $2.2\times 10^{-5}$ & $1.2\times 10^{-6}$ & 0.36 & $6.2\times 10^{-4}$ & $(6.48\pm1.04)\times 10^{-4}$ \\
\hline
$D^0\to K^{*+} \rho^-$ & $1.9\times 10^{-4}$ & $1.7\times 10^{-5}$ & $3.0\times 10^{-6}$ & 0.42 & $2.1\times 10^{-4}$& $$ \\
$D^0\to K^{*0} \rho^0$ & $5.0\times 10^{-5}$ & $4.3\times 10^{-6}$ & $7.6\times 10^{-7}$ & 0.42 & $5.5\times 10^{-5}$ & $$ \\
$D^0\to K^{*0} \omega$ & $4.6\times 10^{-5}$ & $3.8\times 10^{-6}$ & $5.7\times 10^{-7}$ & 0.41 & $5.0\times 10^{-5}$ & $$ \\
\end{tabular}
\end{ruledtabular}}
\end{table}

In the factorization framework, we find that $|H_-|^2\gsim |H_0|^2> |H_+|^2$,~~ $|A_\parallel|^2 \gsim |A_0|^2>|A_\bot|^2$  and $|S|^2>|P|^2>|D|^2$.  Therefore, the longitudinal polarization $f_L$
is expected to be in the vicinity of 0.5 or smaller.  Indeed, $f_L=0.475\pm0.271$ was found by Mark~III in $D^0\to K^{*-}\rho^+$~\cite{MarkIII:VV}.  This is not the case in tree-dominated charmful or charmless $B \to VV$ decays where the longitudinal polarization dominates, i.e., $|H_0|^2> |H_-|^2> |H_+|^2$ and $f_L=1-{\cal O}(m^2_V/ m^2_B)$.  However, for the $D^0\to \ov K^{*0}\rho^0$ decay, it was found by Mark~III~\cite{MarkIII:VV} that this mode proceeded through the transverse polarization, with only a tiny room for the longitudinal polarization.  More precisely, the transverse branching fraction $\B(D^0\to\ov K^{*0}\rho^0)_{T}=(1.6\pm0.6)\%$, while the total branching fraction is $\B(D^0\to\ov K^{*0}\rho^0)_{\rm tot}=(1.59\pm0.35)\%$.  Mark~III also measured the partial-wave branching fractions: $(3.1\pm0.6)\%$, $<3\times 10^{-3}$ and $(2.1\pm0.6)\%$ for the $S$-, $P$- and $D$-waves, respectively~\cite{MarkIII:VV}.

Experimental results for the branching fractions of $D^0\to V\!V$ in partial waves are summarized in Table~\ref{tab:DVVdata}. We notice that all the available measurements of $D_s^+$ and $D^+$ decays to $V\!V$ are performed by BESIII. From the viewpoint of the factorization approach, there exist several puzzles with regard to the data: (i) While one expects $|S|^2>|P|^2>|D|^2$ from na{\"i}ve factorization, $D^0\to K^{*-}\rho^+, \ov K^{*0}\rho^0, \rho^+\rho^-, \rho^0\rho^0$ seem to be dominated by the $D$-wave and $D_s^+\to K^{*0}\rho^+,K^{*-}\rho^0$ are dominated by the $P$-wave. In particular, the $D$-wave dominance is entirely unexpected. (ii) The decay $D^0\to \omega\phi$ is observed by BESIII to be transversely polarized with $f_L<0.24$ ~\cite{BESIII:D0omegaphi}. (iii) If $D^0\to\omega\phi$ and $D^0\to\rho^0\phi$ proceed only through the internal $W$-emission, their branching fractions and polarizations are expected to be the same. Experimentally, they differ not only in rates but also in the polarization. How do we understand the puzzles with the rates and polarizations for $D^0\to\omega\phi$ and $\rho^0\phi$? One possibility is to consider the final-state rescattering of $D^0\to K^{*+}K^{*-}$ which proceeds through external $W$-emission and $W$-exchnage (see Table~\ref{tab:AmpDtoVV}). It is easily seen that final-state interactions of $D^0\to K^{*+}K^{*-}$ will contribute to both $\omega\phi$ and $\rho^0\phi$ through external $W$-emission, but only to the former through $W$-exchange as advocated in Ref.~\cite{Zhao}. Another approach is to include flavor-singlet contributions $S_h$ unique to both $\omega$ and $\phi$:
\be
A(D^0\to \rho^0\phi)={1\over\sqrt{2}}\lambda_s(C_h+S_h^\phi), \qquad
A(D^0\to \omega\phi)={1\over\sqrt{2}}\lambda_s(C_h+S_h^\omega-S_h^\phi).
\en
It is conceivable that $D^0\to\rho^0\phi$ receives flavor-singlet contribution $S_\phi$, while the contributions $S_\phi$ and $S_\omega$ are essentially canceled out in $D^0\to\omega\phi$. We note in passing that the predicted $\B(D_s^+\to\rho^+\eta')$ is substantially smaller than the experiment and this calls for the flavor-singlet contribution from the $\eta'$.

Branching fractions of $D^0\to VV$ in partial waves calculated in the factorization approach are presented in Table~\ref{tab:DVV}.   Since the $W$-exchange contributions are neglected in na{\"i}ve factorization, no estimate is made for the branching fractions of $K^{*0}\ov K^{*0}$, $\ov K^{*0}\phi$, $K^{*0}\phi$ and $\rho^0\omega$. For the effective Wilson coefficients, we have used $a_1=0.90$ and $a_2=-0.64$\,. Comparing Table~\ref{tab:DVV} with the measured partial-wave rates given in Table~\ref{tab:DVVdata} shows that the predictions based on factorization deviate from the experimental measurements. This indicates the necessity of taking into account the non-factorizable $W$-exchange contributions which might account for the $D$-wave dominance observed in  $D^0\to K^{*-}\rho^+, \ov K^{*0}\rho^0, \rho^+\rho^-$ and $\rho^0\rho^0$ channels, an issue to be investigated in the near future. The predicted longitudinal polarization $f_L$ ranges from 0.31 to 0.49.

\section{$D^0$-$\ov D^0$ mixing} \label{sec:y}

The two-body decays $D^0\to PP$ and quasi-two-body decays $D^0\to V\!P, V\!V, SP, SV, AP, AV$, $TP,TV$  account for about 3/4 of the total hadronic rates.
Many of the 3-body final states arise from $SP$, $V\!P$, and $T\!P$ decays, the 4-body states from $V\!V$ and $AP$ decays, and the 5-body states from $AV$ decays. The nonresonant 3-body and 4-body decays are at most 10\% of the multi-body decay rates. Hence, it is arguable that these two-body and quasi-two-body channels dominate and can provide a good estimate of the mixing parameters. As mentioned in the Introduction, use of the TDA is made to reduce the uncertainties with the measured channels and estimate those modes yet to be observed. In particular, we focus on $D\to PP$ and $V\!P$ decays and present updated topological amplitudes.

The general expression for the $D$ mixing parameter $y$ is given by~\cite{Falk:y}
\be \label{eq:y}
 y= \sum_n \eta_{\rm CKM}(n)\eta_{\rm CP}(n)\cos\delta_n\sqrt{\B(D^0\to n)\B(D^0\to\bar n)},
\en
where $\delta_n$ is the strong phase difference between the $D^0\to n$ and $\bar D^0\to n$ amplitudes and $\eta_{\rm CKM}=(-1)^{n_s}$ with $n_s$ being the number of $s$ and $\bar s$ quarks in the final state. The factor $\eta_{\rm CP}=\pm1$ is well-defined since $|n\ra$ and $|\bar n\ra$  are in the same $SU(3)$ multiplet. Hence, this factor is the same for the entire multiplet.

\subsection{ $PP$}
Since $CP|\pi^0\ra=-|\pi^0\ra$ and likewise for $\eta,\eta'$, we will choose the convention that $CP|K^+\ra=-|K^-\ra$ and $CP|K^0\ra=-|\bar K^0\ra$. Because
\be \label{eq:CP}
CP|M_1M_2\ra=\eta_{\rm CP}(M_1)\eta_{\rm CP}(M_2)(-1)^L|M_1M_2\ra,
\en
it is clear that $\eta_{\rm CP}(PP)=1$ for decays into two pseudoscalar mesons. The parameter $y$ arising from the $P\!P$ states is then
\be \label{eq:yPP}
y_{_{P\!P}} &=&\B(\pi^+\pi^-)+\B(\pi^0\pi^0)+\B(\pi^0\eta)+\B(\pi^0\eta')+\B(\eta\eta)
+\B(\eta\eta')+\B(K^+K^-)+\B(K^0\ov K^0) \non \\
&& -2\cos\delta_{K^-\pi^+}\sqrt{\B(K^-\pi^+)\B(K^+\pi^-)}-2\cos\delta_{\bar K^0\pi^0}\sqrt{\B(\ov K^0\pi^0)\B(K^0\pi^0)} \non \\ && -2\cos\delta_{\bar K^0\eta}\sqrt{\B(\ov K^0\eta)\B(K^0\eta)}-2\cos\delta_{\bar K^0\eta'}\sqrt{\B(\ov K^0\eta')\B(K^0\eta')}.
\en
Minus signs appear in the interference terms between the CF and DCS decay modes owing to the negative  $\eta_{\rm CKM}$ factor.

To see that $y$ vanishes in the SU(3) limit, as noted in the Introduction, the contributions to $y$ from the charged pions and kaons
\be
y_{\pi^\pm,K^\mp}=\B(\pi^+\pi^-)+\B(K^+K^-)-2\sqrt{\B(K^-\pi^+)\B(K^+\pi^-)}
\en
vanish in the U-spin limit, where the strong phase $\delta_{K^-\pi^+}\to 0$ in the same limit. To see the cancellation among the neutral states, we work on the SU(3) singlet $\eta_0$ and octet states $\pi, K,\eta_8$. When the SU(3) symmetry is exact, the octet states have the same masses and $D^0\to K^0\ov K^0$ is prohibited. As shown explicitly in Ref.~\cite{Cheng:2010}, perfect cancellation occurs among the SU(3) neutral octet final states and among the decay modes $\pi^0\eta_0$, $\eta_8\eta_0$, $\ov K^0\eta_0$ and $K^0\eta_0$ involving the SU(3) singlet $\eta_0$. Therefore, $y_{PP}$ indeed vanishes in the SU(3) limit.

To compute the mixing parameter $y$, we need to know the phase $\delta_n$. There are four CF $D^0\to PP$ decays, namely, $K^-\pi^+$, $\ov K^0(\pi^0, \eta,\eta')$ and four DCS modes $K^+\pi^-$, $K^0(\pi^0, \eta,\eta')$. From Table~\ref{tab:DCSPP}, we see that $D^0\to K^0(\pi^0, \eta,\eta')$ and $D^0\to \ov K^0(\pi^0, \eta,\eta')$ (see, e.g., Table~I of Ref.~\cite{Cheng:2010}) have the same strong phases and hence $\cos\delta=1$. For $D^0\to K^-\pi^+$ and $K^+\pi^-$, $\delta_{K^+\pi^-}={\rm arg}[(1.23T+E)/(T+E)]=5.83^\circ$  and hence $\cos\delta_{K^+\pi^-}=0.995$ which is consistent with the current experiment measurement of $0.990\pm0.025$~\cite{PDG}.

From the branching fractions of $D^0\to PP$ modes exhibited in Table~\ref{tab:BRPP}, we obtain
\be
y_{_{P\!P}}=
\begin{cases}
(1.113\pm0.007)\%-(1.058\pm0.006)\%=(0.055\pm 0.009)\% &
{\rm Solution~I,} \cr
 (1.227\pm0.010)\%-(1.061\pm0.006)\%=(0.166\pm 0.012)\% &
{\rm Solution~II,}
\end{cases}
\en
for $\phi=40.4^\circ$, and
\be
y_{_{P\!P}}=
\begin{cases}
(1.154\pm0.007)\%-(1.050\pm0.007)\%=(0.102\pm 0.010)\% &
{\rm Solution~I,} \cr
 (1.283\pm0.009)\%-(1.053\pm0.007)\%=(0.231\pm 0.012)\% &
{\rm Solution~II,}
\end{cases}
\en
for $\phi=43.5^\circ$. The difference between solutions I and II arises from the two SCS modes, $D^0\to \eta\eta$ and $D^0\to \eta\eta'$ (see Table~\ref{tab:BRPP}).  SU(3) symmetry breaking occurs in both the decay amplitudes and in the final-state phase space.  In the previous analysis of Ref.~\cite{Falk:y}, the authors considered only SU(3) violation in the phase space and obtained a negative $y$:
\be
y_{PP,8}=-1.8\times 10^{-4}, \qquad y_{PP,27}=-3.4\times 10^{-5},
\en
with $P\!P$ being an {\bf 8} or {\bf 27} SU(3) representation. Indeed, if we neglect SU(3) violation in the decay amplitudes, we will obtain a negative mixing parameter $y$. The channel $D^0\to K^+K^-$ poses the largest SU(3) symmetry breaking in the decay amplitude.

If we use the experimental measurements as the input and employ the predictions based on the TDA for the yet-to-be-measured DCS modes, namely, $D^0\to K^0\pi^0$, $K^0\eta$ and $K^0\eta'$ (cf. Table~\ref{tab:BRPP}), we find
\be
y_{_{P\!P}}=
(1.131\pm0.030)\%-(1.026\pm0.012)\%=(0.110\pm 0.033)\%,
\en
where use of $\cos_{K^+\pi^-}=0.990\pm0.025$~\cite{PDG} has been made.
This is between the two predictions for $\phi=40.4^\circ$ and close to Solution~I for $\phi=43.5^\circ$. Nevertheless, theoretical predictions have smaller uncertainties. Hence, we conclude that
\be
y_{_{P\!P}}\sim (0.110\pm 0.011)\%.
\en
Recall that $y_{_{P\!P}}=(0.086\pm0.041)\%$ was obtained in Ref. \cite{Cheng:Dmixing} and $(0.100\pm0.019)\%$ in Ref. \cite{Jiang}.

\subsection{$V\!P$}

The neutral vector mesons $\rho^0,\omega,\phi$ are {\it CP}-even eigenstates. It is thus convenient to define $C\!P|V\ra=|\ov V\ra$ for the vector meson in the same $SU(3)$ multiplet. It follows from Eq.~(\ref{eq:CP}) that $\eta_{\rm CP}(V\!P)=+1$ for decays into one vector meson and one pseudoscalar meson. There are more decay modes available for the $V\!P$ final states, namely, $V_1P_2$ and $P_1V_2$. There are a total of 30 channels for $V\!P$ (8 for CF, 14 for SCS, and 8 for DCS), to be contrasted with the 16 $P\!P$ channels. The parameter $y$ arising from the $V\!P$ states is given by
\be
y_{_{V\!P}} = y_{_{V\!P,1}} +y_{_{V\!P,2}},
\en
with
\be \label{eq:yVP1}
\begin{split}
y_{_{V\!P,1}}
=& \B(\pi^0\rho^0)+\B(\pi^0\omega)+\B(\pi^0\phi)+\B(\eta\rho^0)+\B(\eta\omega)+\B(\eta'\rho^0)
+\B(\eta'\phi)+\B(\eta'\omega)
\\
& +2\cos\delta_{\pi^+\rho^-}\sqrt{\B(\pi^+\rho^-)\B(\pi^-\rho^+)}+2\cos\delta_{ K^+K^{*-}}\sqrt{\B(K^{*-}K^+)\B(K^-K^{*+})}
\\
& +2\cos\delta_{K^0\bar K^{*0}}\sqrt{\B( K^0\ov K^{*0})\B(\ov K^0 K^{*0})} ,
\end{split}
\en
and
\be \label{eq:yVP2}
\begin{split}
y_{_{V\!P,2}}
=& -2\cos\delta_{\pi^+K^{*-}}\sqrt{\B(K^{*-}\pi^+)\B(K^{*+}\pi^-)}-2\cos\delta_{K^-\rho^+}
\sqrt{\B(K^{-}\rho^+)\B(K^{+}\rho^-)}
\\
& -2\cos\delta_{\pi^0\bar K^{*0}}\sqrt{\B(\ov K^{*0}\pi^0)\B(K^{*0}\pi^0)}-2\cos\delta_{\bar K^0\rho^0}\sqrt{\B(\ov K^{0}\rho^0)\B(K^{0}\rho^0)}
\\
& -2\cos\delta_{\eta\bar K^{*0}}\sqrt{\B(\ov K^{*0}\eta)\B(K^{*0}\eta)}-2\cos\delta_{\eta' \bar K^{*0}}\sqrt{\B(\ov K^{*0}\eta')\B(K^{*0}\eta')}
\\
& -2\cos\delta_{\bar K^0\omega}\sqrt{\B(\ov K^0\omega)\B(K^0\omega)}-2\cos\delta_{\bar K^0\phi}\sqrt{\B(\ov K^0\phi)\B(K^0\phi)} .
\end{split}
\en
In the SU(3) limit, the cancellation of the ninth and tenth terms of $y_{_{V\!P,1}}$ with the first and second terms of  $y_{_{V\!P,2}}$  is obvious as the relevant strong phases are the same and all the $\sqrt{...}$ terms are proportional to $|(T_V+E_P)(T_P+E_V)|$. To see the cancellation among the neutral states, we work on the SU(3) singlets $\eta_0$, $\phi$ and the octet states $\pi$, $K$, $\eta_8$, $\omega$. For octet neutral states, $y_{_{V\!P}}\propto (E_P-E_V)^2$, which vanishes in the limit of SU(3) symmetry. Because the decay constant of the vector meson $f_V$ typically of order 210~MeV is much larger than $f_P$, many $V\!P$ modes have rates greater than the $PP$ ones.  Moreover, the number of $V\!P$ channels is almost double that of $P\!P$ ones. It is thus na{\"i}vely anticipated that the $V\!P$ mode contributions to $y$ ought to be larger than $y_{P\!P}$.

Taking Tables~\ref{tab:VPCF}, \ref{tab:SCSVP} and \ref{tab:DCSVP} as the input for the branching fractions of $D\to V\!P$, we obtain
\be
y_{_{V\!P}}=
\begin{cases}
(2.739\pm0.043)\%-(2.576\pm0.058)\%=(0.163\pm 0.072)\% &
{\rm (F4),} \\
(2.752\pm0.053)\%-(2.381\pm0.045)\%=(0.371\pm 0.069)\% &
{\rm (F1'),}
\end{cases}
\en
where $\cos\delta_n=1$ is assumed for all the modes.  If we employ the calculated $\cos\delta_n$'s given in Table~\ref{tab:phase} based on the TDA, we get
\be \label{eq:yVP}
y_{_{V\!P}}=
\begin{cases}
(2.735\pm0.043)\%-(2.514\pm0.057)\%=(0.220\pm 0.071)\% &
{\rm (F4),} \\
(2.664\pm0.052)\%-(2.229\pm0.043)\%=(0.435\pm 0.068)\% &
{\rm (F1').}
\end{cases}
\en
The predicted value of $y$ is much larger for solution (F1') for two reasons: (i) the predicted DCS branching fractions using (F1') are smaller compared to that using solution (F4) (see Table~\ref{tab:DCSVP}) due to smaller $E_V''$ and $E_P''$ [cf. Eqs.~(\ref{eq:EVpp}) and (\ref{eq:EPpp})], and (ii) the phase terms $\cos\delta_n$'s in (F1') are smaller in magnitude than that in (F4). Hence, these two features render the second term $y_{_{V\!P,2}}$ smaller in (F1').

It is interesting to notice that if we use the experimental data as the input and employ the predictions based on the TDA for those modes yet to be measured, we get
\be
y_{_{V\!P}}=\begin{cases}
(2.813\pm0.059)\%-(2.547\pm0.184)\%=(0.266\pm 0.193)\% &
{\rm (F4),} \cr
(2.811\pm0.059)\%-(2.365\pm0.183)\%=(0.446\pm 0.192)\% &
{\rm (F1'),}
\end{cases}
\en
for $\cos\delta_n=1$ and
\be
y_{_{V\!P}}=\begin{cases}
(2.808\pm0.059)\%-(2.462\pm0.183)\%=(0.322\pm 0.193)\% &
{\rm (F4),} \cr
(2.721\pm0.058)\%-(2.214\pm0.182)\%=(0.547\pm 0.191)\% &
{\rm (F1'),}
\end{cases}
\en
for  the calculated $\cos\delta_n$ in (F4) and (F1').
Evidently, the uncertainties are substantially reduced in the diagrammatic approach. We thus have the lower bound on $y_{_{V\!P}}$
\be
y_{_{V\!P}}\gsim (0.220\pm0.071)\%,
\en
see Eq. (\ref{eq:yVP}).  We recall that $y_{_{V\!P}}=(0.112\pm0.072)\%$ in Ref. \cite{Jiang}, and
$y_{_{V\!P}}=(0.269\pm0.253)\%$, $(0.152\pm0.220)\%$ in schemes (A,A1) and (S,S1), respectively \cite{Cheng:Dmixing}.

In Ref.~\cite{Falk:y} where the phase space is the only source of SU(3) violation, the values of $y$ for ${\bf 8}_S$, ${\bf 8}_A$, ${\bf 10}$, $\ov {\bf 10}$ and {\bf 27} representations are estimated to be 0.15\%, 0.15\%, 0.10\%, 0.08\% and 0.19\%, respectively. Unlike the previous $P\!P$ case, the sign of $y_{_{V\!P}}$ is positive. Recall that the large rate disparity between $D^0\to K^+K^-$ and
$D^0\to \pi^+\pi^-$ implies large SU(3) breaking effects in the amplitude of $T+E$, more precisely, $|T+E|_{K\!K}/|T+E|_{\pi\pi}\approx 1.80$~\cite{Cheng:2019}. In the $V\!P$ sector, instead
we have $\Gamma(K^+K^{*-})<\Gamma(\pi^+\rho^-)$ and $\Gamma(K^-K^{*+})<\Gamma(\pi^-\rho^+)$. This is understandable as the available phase space is proportional to $p_c^3/m_V^2$ [see  Eq.~(\ref{decaywidthA})], this explains why $\Gamma(D^0\to KK^*)<\Gamma(D^0\to \pi\rho)$ owing to the fact that $p_c(\pi\rho)=764$~MeV and $p_c(KK^*)=608$~MeV. As shown in Ref.~\cite{Cheng:2019}, we find from the measured branching fractions that
\be \label{eq:TEinVP}
{|T_V+E_P|_{\pi^+\rho^-}\over |T_V+E_P|_{K^+K^{*-}} }=1.08\,, \qquad\quad {|T_P+E_V|_{\pi^-\rho^+}\over |T_V+E_P|_{K^-K^{*+}} }=0.91\,.
\en
This implies that SU(3) breaking in the amplitudes of $T_V+E_P$ and $T_P+E_V$ is small, contrary to the $P\!P$ case. This means that SU(3) violation in the decay amplitudes plays a less significant role in $D\to V\!P$ decays. Therefore, an estimate of $y_{_{V\!P}}$ solely based on SU(3) symmetry breaking in the phase space leads to a correct sign of $y$.

\subsection{$V\!V$}

The $VV$ states with different partial waves contribute with different {\it CP} parties. We have $\eta_{\rm CP}(VV)=1$ for $VV$ in the $S$ or $D$ wave, and $-1$ in the $P$ wave~\cite{Falk:y}. The parameter $y$ for $VV$ modes has the expression
\be \label{eq:yVV}
\begin{split}
y_{_{VV,\ell}} =&\B(\rho^+\rho^-)_\ell+\B(\rho^0\rho^0)_\ell+\B(\rho^0\omega)_\ell+\B(\rho^0\phi)_\ell
+\B(\omega\omega)_\ell +\B(\omega\phi)_\ell +\B(\phi\phi)_\ell
\\
&+\B(K^{*+}K^{*-})_\ell + \B(K^{*0}\bar K^{*0})_\ell -2\cos\delta_{K^{*-}\rho^+}\sqrt{\B(K^{*-}\rho^+)_\ell\B(K^{*+}\rho^-)_\ell}
\\
&-2\cos\delta_{\bar K^{*0}\rho^0}\sqrt{\B(\ov K^{*0}\rho^0)_\ell\B(K^{*0}\rho^0)_\ell}
- 2\cos\delta_{\bar K^{*0}\omega}\sqrt{\B(\ov K^{*0}\omega)_\ell\B(K^{*0}\omega)_\ell}
\\
&
-2\cos\delta_{\bar K^{*0}\phi}\sqrt{\B(\ov K^{*0}\phi)_\ell\B(K^{*0}\phi)_\ell}
\end{split}
\en
for $\ell=S,D$, and an overall minus sign is needed for $\ell=P$. From Table~\ref{tab:AmpDtoVV}, it is easily seen that the contribution
\be
y_{K^*,\rho} = \B(\rho^+\rho^-)_\ell+\B(K^{*+}K^{*-})_\ell-2\cos\delta_{K^{*-}\rho^+}
\sqrt{\B(K^{*-}\rho^+)_\ell\B(K^{*+}\rho^-)_\ell}
\en
vanishes in the SU(3) limit. This $U$-spin relation cannot be tested by the current data as only the $D^0\to K^{*-}\rho^+$ and $D^0\to \rho^+\rho^-$ decays have been measured.  Notice that although $D^0\to \phi\phi$ is prohibited by phase space, it is needed in order to show a vanishing $y_{_{VV,\ell}}$ in the SU(3) limit.

An estimate using the na{\"i}ve factorization results from Table~\ref{tab:DVV} yields
\be
y_{_{V\!V,S}}=-0.167\%, \qquad y_{_{V\!V,P}}=1.61\times 10^{-4}, \qquad y_{_{V\!V,D}}=4.71\times 10^{-5}.
\en
Since the predicted rates of $D^0\to VV$ in partial waves based on na{\"i}ve factorization do not resemble the data given in Table~\ref{tab:DVVdata}, particularly for the $D$-wave dominance in the $D^0\to K^{*-}\rho^+$, $\ov K^{*0}\rho^0$ , $\rho^+\rho^-$ and $\rho^0\rho^0$ decays as suggested by the current data, it is premature to have a reliable estimate of $y_{_{V\!V,\ell}}$. Indeed, if we use the data in Table~\ref{tab:DVVdata} as the input and employ the predictions based on na{\"i}ve factorization for those modes yet to be measured, we get
\be
y_{_{V\!V,S}}=0.0271\%, \qquad y_{_{V\!V,P}}=-0.0958\%, \qquad y_{_{V\!V,D}}=0.361\%.
\en

The results obtained in Ref.~\cite{Falk:y} in which only SU(3) violation in the phase space is considered are given by
\be
\begin{split}
&S\mbox{-wave}: \qquad y_{_{V\!V,8}}=-0.39\%, \qquad y_{_{V\!V,27}}=-0.30\%,
\\
&P\mbox{-wave}: \qquad y_{_{V\!V,8}}=-0.48\%,  \qquad y_{_{V\!V,27}}=-0.70\%,
\\
&D\mbox{-wave}: \qquad y_{_{V\!V,8}}=2.5\%,  \qquad\quad y_{_{V\!V,27}}=2.8\%.
\end{split}
\en
Because of the momentum dependence of the $D$-wave is proportional to $p_c^3$, the $D$-wave phase space is most sensitive to the SU(3) breaking in $p_c$.

\subsection{$D\to (S,A,T)(P,V)$ decays}

There are hadronic $D$ decays into an even-parity meson $M$ and a pseudoscalar meson or a vector meson, where $M$ represents a scalar meson $S$, an axial-vector meson $A$, or a tensor meson $T$. They have been studied in the literature~\cite{Cheng:2010SAT,Cheng:2022SP,Cheng:2022TP}, but the data are not adequate to allow for the extraction of topological amplitudes, especially for the $W$-exchange amplitudes. These will have to be left to a future investigation.

\section{Conclusions \label{sec:Conclusions}}

In this work, we have presented an updated analysis of the two-body decays $D\to PP$, $V\!P$ and $VV$ decays within the framework of the topological diagram approach. For the CF decay modes involving a neutral kaon, $K_S^0$ or $K_L^0$, the relation $\Gamma(\ov K^0)=2\Gamma(K_S^0)$ can be invalidated by the interference between the CF and DCS amplitudes. Since the topological amplitudes in DCS modes are not necessarily the same as those in CF ones beyond flavor SU(3) symmetry, we prefer to use the good approximate relation $\B(D\to \ov K^0 M)\cong \B(D\to K_S^0 M)+\B(D\to K_L^0 M)$ for $M=P$ and $V$.

In the $PP$ sector, the tree topological amplitudes $T$, $C$, $E$ and $A$ are extracted from CF $D\to PP$ decays for the $\eta-\eta'$ mixing angle $\phi = 40.4^\circ$ and $43.5^\circ$, respectively. The fitted $\chi^2$ values almost vanish with the quality close to unity for $\phi=40.4^\circ$.

Assuming that the double-primed amplitudes in the DCS sector are the same as the unprimed ones, the calculated $K_S^0-K_L^0$ asymmetries $R(D^0,P)$ for $D^0\to K_{S,L}^0P$ decays with $P=\pi^0$, $\eta$ and $\eta'$ agree with the experiment, meaning that $D^0\to K_S^0P$ has a rate larger than that of $D^0\to K_L^0P$. However, our predicted $R(D^+,\pi^+)$ is opposite to the experiment in sign and the calculated $R(D_s^+, K^+)$ is too small compared to the data. This is ascribed to the fact that the relative phase between $(C+A)$ and $(T+C)$ is slightly larger than $90^\circ$, rendering the interference between $D^+\to \ov K^0\pi^+$ and $D^+\to K^0\pi^+$ destructive in $D^+\to K_S^0\pi^+$ and constructive in $D^+\to K_L^0\pi^+$. This is opposite to the pattern observed experimentally. We find that if the phase difference is decreased slightly by $10^\circ$, we are able to accommodate both
$R(D^+,\pi^+)$ and $R(D_s^+, K^+)$.

In the $V\!P$ sector, if the double-primed topological amplitudes in DCS decays are taken to be the same as the unprimed ones, we will be led to some predictions not in accordance with the experiment. That is why we prefer to apply the relation $\B(D\to \ov K^0 V)\cong \B(D\to K_S^0 V)+\B(D\to K_L^0 V)$. Unfortunately, we have the data of $\B(D\to K_L^0 V)$  for $V=\omega$ and $\phi$, but not for $V=\rho^0$, $\rho^+$ and $K^{*+}$. Since it is most likely that $E_{V,P}''\neq E_{V,P}$, we fit the topological amplitudes to $D^0\to \ov K^0(\omega,\phi)$, $D^+\to K_S^0\rho^+$ and $D_s^+\to K_S^0 K^{*+}$, but not $D^0\to K_S^0\rho^0$.

A global fit to the CF modes in the $V\!P$ sector gives many solutions with similarly small local minima in $\chi^2$: (F1)-(F5) for $\phi=40.4^\circ$ and (F1')-(F5') for $\phi=43.5^\circ$, when we restrict ourselves to $\chi^2_{\rm min} < 10$. The solution degeneracy is lifted once we use them to predict for the  SCS modes.  In the end, we find that only Solutions (F4) and (F1') can accommodate all SCS modes except that the predicted $\B(D^+\to \pi^+\omega)$ in (F1') is slightly larger than the data. These two solutions have a common feature that $C_V$ and $C_P$ are close in phase in order to simultaneously explain the small ${\cal B}(D^0 \to \pi^0\omega)$ and large ${\cal B}(D^0 \to \pi^0\rho^0)$.  The annihilation amplitudes $A_V$ and $A_P$ are more precisely determined than before because of a significantly improved new measurement of $D_s^+\to \pi^+\rho^0$; they are comparable in size and similar in phase.

For DCS $D\to V\!P$ decays, the assumption of $E_{V,P}''=E_{V,P}$ leads to some inconsistencies with the experiment; for example, the predicted strong phases $\delta_{K^{*+}\pi^-}$ and $\delta_{K^{*0}\pi^0}$ are too large compared to their experimental values. The relations of $E_{V,P}''$ with $E_{V,P}$ are given in Eqs.~(\ref{eq:EVpp}) and (\ref{eq:EPpp}), respectively.

The $K_S^0-K_L^0$ asymmetries in the $D\to K_{S,L}^0V$ decays are shown in Table~\ref{tab:VPKLKS}. The calculated $R(D^0,\omega)$ and $R(D^0,\phi)$ do not agree with the experiment. We conjecture that additional singlet contributions due to SU(3)-singlet nature of $\omega$ and $\phi$ should account for the discrepancy.

Thanks to BESIII, many new data on $D_s^+$ and $D^+$ to $V\!V$ decays became available in recent years. In the meantime, several new puzzles have also emerged. For example, $D^0\to K^{*-}\rho^+$, $\ov K^{*0}\rho^0$, $\rho^+\rho^-$, and $\rho^0\rho^0$ seem to be dominated by the $D$-wave, while $D_s^+\to K^{*0}\rho^+$ and $K^{*-}\rho^0$ dominated by the $P$-wave, contrary to the na{\"i}ve expectation of $S$-wave dominance. In spite of the progresses in the field, the data are still not adequate to allow for a meaningful extraction of helicity or partial-wave amplitudes, especially for the $W$-exchange and $W$-annihilation ones.

Yet another goal of this analysis is to evaluate the $D^0$-$\ov D^0$ mixing parameter $y$ using the exclusive approach through the 2-body decays of the $D^0$ meson. The mixing parameter is usually small owing to large cancellation between the SCS terms and the interference of CF and DCS terms.
As the topological amplitude analysis is available for $D\to PP$ and $V\!P$ decays, we are able to estimate $y_{_{P\!P}}$ and $y_{_{V\!P}}$ more reliably. We conclude that $y_{_{P\!P}}\sim (0.110\pm 0.011)\%$ and the lower bound on $y_{_{V\!P}}$ is $(0.220\pm 0.071)\%$. It is thus conceivable that at least half of the $D^0$-$\ov D^0$ mixing parameter $y$ is accounted for by the $PP$ and $V\!P$ modes. The main uncertainties arise from the DCS channels yet to be measured and the phase factors $\cos\delta_n$'s.

\section*{Acknowledgments}

This research was supported in part by the Ministry of Science and Technology of R.O.C. under Grant Nos.~MOST-107-2119-M-001-034 and MOST-108-2112-M-002-005-MY3.


\end{document}